\documentclass[aps
,pra
,amsmath
,amssymb
,amsfonts
,twocolumn
,letterpaper
,10pt
,tightenlines
,superscriptaddress
,nofootinbib
,nobalancelastpage
,eqsecnum
,longbibliography
,notitlepage
]{revtex4-2}

\usepackage{nag}
\usepackage{graphicx}
\usepackage{amsthm}
\usepackage{tabulary}
\usepackage{qcircuit}
\usepackage{mathtools}
\usepackage{bm}
\usepackage{xcolor}
\usepackage{braket}
\usepackage{datetime}
\usepackage{stackrel}
\usepackage{stmaryrd}
\usepackage{dsfont}
\usepackage{qcircuit}
\usepackage[english]{babel}
\usepackage{units}

\usepackage{tikz}
\usetikzlibrary{backgrounds,decorations.pathreplacing}

\usepackage{chngcntr}
\counterwithout{equation}{section}

\usepackage[normalem]{ulem}
\usepackage{slashed}
\usepackage{soul}

\usepackage{xcolor}
\usepackage[
    colorlinks=true,
    urlcolor=blue,
    linkcolor=blue,
    citecolor=blue,
    filecolor=blue,
]{hyperref}
\usepackage[caption=false]{subfig}
\usepackage{amssymb}
\usepackage{amsmath}
\usepackage{amsthm}
\usepackage{braket}

\usepackage[OT2,T1]{fontenc}
\DeclareSymbolFont{cyrletters}{OT2}{wncyr}{m}{n}
\DeclareMathSymbol{\Sha}{\mathalpha}{cyrletters}{"58}

\usepackage{makeidx}
\makeindex

\usepackage{soul}
\usepackage{xargs}
\usepackage{todonotes}
\newcommandx{\cmnote}[2][1=]{\linespread{1.0}\todo[linecolor=red,backgroundcolor=red!25,bordercolor=red,#1]{#2}}

\let\underline\ul

\allowdisplaybreaks[4]

\DeclareMathOperator{\Tr}{Tr}

\renewcommand{\Re}{\operatorname{Re}}
\renewcommand{\Im}{\operatorname{Im}}

\DeclareMathOperator{\sech}{sech}

 \index{} \index{} \index{} \index{} \index{} \index{} \index{} \index{}
\makeatletter

\newcommand{\ringplus}{\mathbin{\text{\@ringplus}}}

\newcommand{\@ringplus}{%
  \ooalign{\hidewidth\raise1.3ex\hbox{\tiny$\circ$}\hidewidth\cr$\m@th+$\cr}%
}

\newcommand{\ringminus}{\mathbin{\text{\@ringminus}}}

\newcommand{\@ringminus}{%
  \ooalign{\hidewidth\raise0.9ex\hbox{\tiny$\circ$}\hidewidth\cr$\m@th-$\cr}%
}
\makeatother
 \index{} \index{} \index{} \index{} \index{} \index{} \index{} \index{}

\newcommand{\tp}[0]{\mathrm{T}}

\newcommand{\EPR}{\text{EPR}}
\newcommand{\bounceEPR}[2]{\op{A}_{#2}(#1)}
\newcommand{\bounceEPRgate}[2]{A_{#2}(#1)}

\newcommand{\dampop}[1]{\op{N}(#1)}
\newcommand{\dampgate}{N(\beta)}

\newcommand{\GKP}{\text{GKP}} 
\newcommand{\GKPproj}{\op{\Pi}_\GKP}
\newcommand{\approxGKPproj}{\hat{\overline{\Pi}}_\GKP}
\newcommand{\qunaught}{\varnothing}

\DeclareFontFamily{U}{wncy}{}
\DeclareFontShape{U}{wncy}{m}{n}{<->wncyr10}{}
\DeclareSymbolFont{mcy}{U}{wncy}{m}{n}
\DeclareMathSymbol{\Sh}{\mathord}{mcy}{"58}

\newcommand{\negspace}{\!}

\newcommand{\lsub}[2]{{\protect\vphantom{#1}}_{#2} \negspace {#1}}
\newcommand{\rsub}[2]{{#1} \negspace {\protect\vphantom{#1}}_{#2}}
\newcommand{\lrsub}[3]{{\protect\vphantom{#1}}_{#2} \negspace {#1} \negspace {\protect\vphantom{#1}}_{#3}}

\newcommand{\ketsub}[2]{\rsub {\ket{#1}} {#2}}
\newcommand{\brasub}[2]{\lsub {\bra{#1}} {#2}}

\newcommand{\qbra}[1]{\brasub{#1} q}
\newcommand{\pket}[1]{\ketsub{#1} p}
\newcommand{\qket}[1]{\ketsub{#1} q}

\newcommand{\inprod}[2]{\left\langle {#1} | {#2} \right\rangle}

\newcommand{\inprodsubsub}[4]{\lrsub {\inprod{#1}{#2}} {#3} {#4}}

\newcommand{\outprod}[2]{\ket {#1}\!\bra {#2}}

\newcommand{\outprodsubsub}[4]{\ketsub {#1}{#3} \brasub{#2}{#4}}

\newcommand{\qoutprod}[2]{\outprodsubsub{#1}{#2}q q}

\newcommand{\avg}[1]{\left\langle {#1} \right\rangle}

\newcommand{\abs}[1]{\left\lvert{#1}\right\rvert}

\newcommand{\op}[1]{\hat{#1}}

\newcommand{\mat}[1]{\bm{\mathrm{#1}}}

 \index{} \index{} \index{} \index{} \index{} \index{} \index{} \index{}
 \index{} \index{} \index{} \index{} \index{} \index{} \index{} \index{}

\newcommand{\blk}{\color{black}}

 \index{} \index{} \index{} \index{} \index{} \index{} \index{} \index{}
 \index{} \index{} \index{} \index{} \index{} \index{} \index{} \index{}
\usepackage{graphicx}

\newcommand{\bsop}{\op{B} }

\usepackage{graphicx}

\newcommand{\qwax}[1][-1]{\ar @{->} [#1,0]}
\newcommand{\bsbal}[1]{\qwax[{#1}] \qw}

\begin{document}

\title{Continuous-variable gate teleportation and bosonic-code error correction}

\newpage
\setcounter{page}{1}
\pagenumbering{arabic}

\author{Blayney W. Walshe}
\email{blayneyw@gmail.com}
\affiliation{Centre for Quantum Computation and Communication Technology, School of Science, RMIT University, Melbourne, VIC 3000, Australia}
\author{Ben Q. Baragiola}
\affiliation{Centre for Quantum Computation and Communication Technology, School of Science, RMIT University, Melbourne, VIC 3000, Australia}
\author{Rafael N. Alexander}
\affiliation{Centre for Quantum Computation and Communication Technology, School of Science, RMIT University, Melbourne, VIC 3000, Australia}
\affiliation{Center for Quantum Information and Control, University of New Mexico, Albuquerque, NM 87131, USA}
\author{Nicolas C. Menicucci}
\affiliation{Centre for Quantum Computation and Communication Technology, School of Science, RMIT University, Melbourne, VIC 3000, Australia}
 \begin{abstract}
We examine continuous-variable gate teleportation using entangled states made from pure product states sent through a beam splitter. 
We show that such states are Choi states for a (typically) non-unitary gate, and we derive the associated Kraus operator for teleportation, which can be used to realize non-Gaussian, non-unitary quantum operations on an input state.
With this result, we show how gate teleportation is used to perform error correction on bosonic qubits encoded using the Gottesman-Kitaev-Preskill code.
This result is presented in the context of deterministically produced macronode cluster states, generated by constant-depth linear optical networks, supplemented with a probabilistic supply of GKP states. 
The upshot of our technique is that state injection for both gate teleportation and error correction can be achieved without active squeezing operations---an experimental bottleneck for quantum optical implementations.

\end{abstract}
\maketitle

\section{Introduction}
\label{scrivauto:12}

Recent strides in the experimental generation of continuous-variable (CV) cluster states \cite{Yokoyama2013, Asavanant2019, Larsen2019} prove that the use of CV measurement-based quantum computing (MBQC) is one of the most promising methods of achieving fault-tolerant universal quantum computing. MBQC utilises a highly entangled resource---known as a cluster state---as its substrate for quantum computing \cite{Raussendorf2001} and only requires adaptive local measurements on the cluster state to implement quantum gates. 

Constructing cluster states using bosonic modes has the advantage of deterministic entanglement generation using linear optics and is highly scalable \cite{Menicucci2011a}.
Unfortunately, computation with CV cluster states is burdened by  intrinsic noise due to finite energy constraints~\cite{Alexander2014}.
Despite this, CV MBQC on a cluster state generated from squeezed states meeting a threshold of 15-17 dB of measured squeezing 
is fault-tolerant \cite{Nick2014}---provided it is supplemented with a source of high-quality bosonic qubits.
Furthermore, this squeezing threshold is not affected by decoherence that manifests as anti-squeezing~\cite{Walshe2019}.

CV cluster states are multi-mode Gaussian states of light specified by a complex-weighted graph~\cite{Menicucci2011}. The states originally considered in Refs.~\cite{Menicucci2006, Gu2009} possessed simple graphs (\emph{e.g.}, a 2D square lattice); however, known methods for their generation require either inline squeezing (active transformations on states other than the vacuum) or a linear optics network that grows with the system size. Related states with a multi-layered graph structure can be generated in a more experimentally feasible way with circuits consisting of \emph{offline squeezing and local constant depth linear optics}~\cite{Menicucci2008, flammia2009optical, Menicucci2011a, Wang2014, Alexander2016, Alexander2018, Larsen2019, Asavanant2019, wu2020quantum}.
 Such states have been generated on a large scale, both in one \cite{Yokoyama2013, Chen2014, Yoshikawa2016} and two dimensions \cite{Larsen2019, Asavanant2019}.
 MBQC on these multi-layered states respects the tensor product structure of \emph{macronodes}, with physical lattice sites made up of one mode from each layer~\cite{Alexander2014, alexander2016flexible}.

The fundamental primitive of CV MBQC schemes with multi-layered graphs is CV teleportation~\cite{lloyd1999quantum, furusawa1998unconditional}. Gaussian homodyne detection teleports quantum information from one node to another with a fidelity depending on the quality of the shared entanglement between the nodes. Ideally, nodes share a maximally entangled EPR state, which allows for perfect teleporation. Up to local phase delays, macronode continuous-variable cluster states (CVCSs) are indeed just a collection of approximate EPR states stitched together at macronodes by non-local measurements (more specifically, a sequence of 50:50 beam splitters followed by homodyne detection on all modes). Gaussian CV MBQC schemes leverage the choice of rotated homodyne bases as degrees of freedom to realize Gaussian operations on the teleported state. This was recently demonstrated in Ref.~\cite{Asavanant2020}.
 
\begin{figure}[t!]
\centering
\includegraphics[width=1\columnwidth]{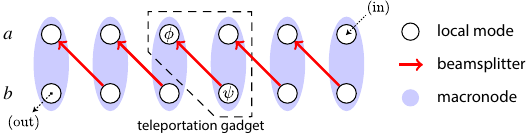}
\caption{
Schematic of a macronode wire, which is a chain of maximally entangled pairs (modes connected by beam splitters) that meet at two-mode sites called \emph{macronodes}. 
Non-local measurements of each macronode, achievable by an additional beam splitter and local measurements at each macronode (not indicated), teleport an input state along the macronode wire (from right to left). In this work, we replace each entangled pair with a \emph{Kraus state} consisting of two arbitrary states, $\ket{\psi}$ and $\ket{\phi}$, coupled on a beam splitter. This realizes a teleported CV gate, which can be non-unitary, depending on the Kraus state.  
A single instance of this CV gate teleportation, which we call the \emph{teleportation gadget}, is enclosed in the dashed box. The macronode-wire circuit diagram for gate teleportation is given in Fig.~\ref{macronodecircuit}.
}\label{macronodecartoon}
\end{figure}

In this work, we provide a method to go beyond Gaussian operations by replacing each entangled pair with a more general state---two arbitrary pure states coupled on a beam splitter, see Fig.~\ref{macronodecartoon}. If either (or both) of these states is non-Gaussian, teleportation can realize a non-Gaussian gate, extending multi-mode Gaussian resources to universal resources for quantum computation.  
Our approach runs parallel to and extends the standard description of gate teleportation~\cite{Gottesman1999, Loock2007, Ewert2019}, where a unitary gate to be teleported is applied to one half of a maximally entangled state. Using the entangled-state method here additionally allows teleportation of non-unitary operations, such as projections, which are useful for bosonic-code error correction. This approach is compatible with a wide variety of non-Gaussian resources~\cite{miyata2016implementation, marshall2015repeat, arzani2017polynomial, walschaers2018tailoring} and does not require active squeezing operations, making it convenient for conventional quantum optics setups.

We apply our analysis to a particular example: gate-teleported error correction of the  Gottesman-Kitaev-Preskill (GKP) code. GKP states allow encoding of digital quantum information in the continuous Hilbert space of a CV mode~\cite{GKP}. Remarkably, this single non-Gaussian resource extends multi-mode Gaussian computation to universal and fault-tolerant quantum computation~\cite{Baragiola2019,Yamasaki2020}. 

GKP states have recently been realized in trapped ion systems~\cite{Fluhmann2019} and superconducting qubit architectures \cite{Campagne-Ibarcq2020}, and multiple novel approaches have been proposed for their generation in optics \cite{Travaglione2002,  Pirandola2004, Vasconcelos2010, Su2019, Eaton2019, Motes2017, Tzitrin2020}. We show that teleporting through entangled pairs made from a probabilistic source of GKP grid states implements GKP error correction.

 Thus, the entire scheme for universal and fault-tolerant quantum computing consists only of offline preparation of GKP and squeezed vacuum states, constant-depth linear optics, and homodyne detection. This is an improvement over prior proposals that require active transformations to combine GKP states with CV cluster states~\cite{Nick2014, Alexander2018, larsen2020architecture}. Finally, we note that our analysis extends to higher dimensional CV cluster states that have two- and three-dimensional structure.

\section{Macronode-based CV Cluster States}
\label{macronode}
A continuous-variable cluster state (CVCS) is a large, Gaussian state across many modes that can serve as a resource for CV teleportation \cite{Menicucci2006} and measurement-based quantum computing (MBQC) \cite{Nick2014}. Canonical CVCSs are constructed using $e^{i g \op{q} \otimes \op{q}}$ interactions to successively entangle momentum eigenstates, and Gaussian homodyne measurements of the modes are used to teleport and enact Gaussian operations on encoded information \cite{Menicucci2006}. An alternative construction employing constant-depth %
passive elements (beam splitters and phase delays) and finitely squeezed states, called a macronode CVCS, provides a blueprint for experimental generation of a CVCS \cite{Alexander2014} that makes better use of the fundamental quantum resource in such states: squeezing \cite{Braunstein2005, van2007building}. Here, two-mode squeezed states are generated using beam splitters; these entangled pairs can then be coupled to others using more beam splitters to create macronode CVCSs with more complicated structures. In this construction, certain collections of physical modes are grouped into \emph{macronodes}, across which logical encoded information is distributed nonlocally. Processing the information within a macronode requires homodyne measurement of its constituent physical modes.

The smallest nontrivial macronode CVCS, the \textit{macronode wire} (see Figure \ref{macronodecartoon}), has two modes per macronode. Larger and more complex macronode CVCSs contain more modes per macronode \cite{Asavanant2019, Larsen2019, wu2020quantum} but can be viewed quantitatively as tools to route information along configurable macronode wires \cite{alexander2016flexible}. Macronode wires have been generated experimentally from both temporal and spectral modes~\cite{Yokoyama2013, Chen2014, Yoshikawa2016} and recently adaptive homodyne measurements have been used to implement Gaussian operations on a temporal-mode macronode wire \cite{Asavanant2020}.

\subsection{Structure of a macronode wire}
A macronode-based CV cluster state~\cite{flammia2009optical} differs from a canonical one~\cite{Menicucci2006,Gu2009} by the fact that multiple modes occupy one logical site within the overarching graph representing the entanglement structure of the state. (For further details, see Ref.~\cite{flammia2009optical}.) These modes are logically grouped together and referred to as a \emph{macronode} to emphasise that it comprises multiple ``micronodes'', each of which is an individual mode. All of the experiments demonstrating large-scale continuous-variable cluster states have generated macronode-based cluster states~\cite{Chen2014,Yokoyama2013,Yoshikawa2016,Asavanant2019,Larsen2019,Asavanant2020}.

We use the following notation to describe the individual modes within a macronode-based CV cluster state. For each local mode~$a$, we define the position and momentum quadratures, $\op{q} = \frac{1}{\sqrt{2}}(\op{a} + \op{a}^\dagger)$ and $\op{p} = \frac{-i}{\sqrt{2}}(\op{a} - \op{a}^\dagger )$, respectively, satisfying $[\op{q}, \op{p}] = i$. This means that the vacuum variance in both quadratures of any mode is~$\avg{\op q^2}_{\text{vac}} = \avg{\op p^2}_{\text{vac}} = 1/2$, which can be interpreted as equivalent to the convention that $\hbar = 1$.

Each quadrature has an associated set of eigenstates, $\qket{s}$ and $\pket{t}$, satisfying $\op{q} \qket{s} = s \qket{s}$ and $\op{p} \pket{t} = t \pket{t}$. 
Information in a one-dimensional macronode-based CV cluster state is encoded in the distributed symmetric ($+$) and antisymmetric ($-$) subspace of the two local modes in a single macronode, arising from a change in the tensor-product structure due to the beam splitter transformation coupling the two modes prior to the final measurements~\cite{Alexander2014}. The balanced beam splitter convention we use here,
    \begin{equation} \label{beamsplitterdef}
        \bsop_{jk} \coloneqq e^{-i \frac{\pi}{4}(\op{q}_j \otimes \op{p}_k - \op{p}_j \otimes \op{q}_k )},
    \end{equation}
generates the distributed quadrature operators at each macronode:
    \begin{align} \label{bstransformation}
        \hat q_{\pm} &\coloneqq 
        \tfrac{1}{\sqrt{2}} (\hat q_{a} \pm \hat q_{b}) \\
        \hat p_{\pm} &\coloneqq \tfrac{1}{\sqrt{2}} (\hat p_{a} \pm \hat p_{b}).
    \end{align}
Specifically,
$ \bsop^\dagger_{ab} (\hat q_a, \hat q_b, \hat p_a, \hat p_b) \bsop_{ab} = (\hat q_-, \hat q_+, \hat p_-, \hat p_+)$, respectively.
Note that the beam splitter we use is not symmetric; $\bsop_{jk}^\dagger = \bsop_{kj} \neq \bsop_{jk}$. In circuit diagrams (which we use extensively below), we represent the beam splitter as an arrow pointing from mode $j$ to mode $k$:
\begin{equation}\label{BScircuit}
\begin{split}
    \Qcircuit @C=1.25em @R=2.1em {
	 &                                                                                  && \bsbal{1} & \rstick{j} \qw  \\
	 & \lstick{\raisebox{2.9em}{$\bsop_{jk} =$}}&& \qw       & \rstick{k} \qw \\
		}
\end{split}		
\end{equation}
A macronode wire is a chain of entangled pairs that are logically linked together at macronodes, a grouping of local modes. Non-local measurements on the macronodes, a beam splitter interaction followed by local measurements at each local modes, implements a string of sequential teleportations (from right to left), see Fig.~\ref{macronodecartoon}.
For more information, see Ref.~\cite{Menicucci2011a, Alexander2014}.

\subsection{Measurement-based CV computation with a macronode CVCS wire\label{sec:MBQCmacronode}} 

In an ideal setting, measurement-based computing with a macronode CVCS begins by preparing every local mode in either a position or a momentum eigenstate, $\qket{0}$ or  $\pket{0}$. Pairs of adjacent modes are first coupled with a beam splitter to generate infinitely squeezed two-mode squeezed states of the form:
\begin{equation}\label{scrivauto:771}
\begin{split}
    \Qcircuit @C=1.25em @R=2.25em {
	 &                                                                                  && \bsbal{1} & \rstick{\ketsub{0}{p_a}} \qw  \\
	 & \lstick{\raisebox{2.9em}{$\bsop_{ab} \ketsub{0}{p_a} \otimes \ketsub{0}{q_b} =$}}&& \qw       & \rstick{\ketsub{0}{q_b}} \qw \\
		}
\end{split}		
\end{equation}
In physical settings, quadrature eigenstates are replaced by their finite-energy, squeezed approximations, which will be discussed in Sec.~\ref{sec:dampedancillae}.

Homodyne measurements of a macronode's two constituent local modes (followed by outcome-dependent displacements) teleport a state encoded in the symmetric mode of that macronode to the symmetric mode of the next macronode. 
The macronode wire's utility for quantum information processing arises from the fact that measuring in rotated bases over two successive macronodes additionally implements any single-mode Gaussian unitary gate \cite{Alexander2014}. The essentials of this procedure are outlined below. 

We define a rotated momentum quadrature, 
    \begin{align}
        \op{p}_{\theta} &\coloneqq \op{R}^\dagger(\theta) \op{p} \op{R}(\theta) =  \op q \sin \theta + \op p \cos \theta \, ,
    \end{align}
where the phase-delay operator 
    \begin{equation} \label{phasedelay}
        \op R(\theta) \coloneqq e^{i \theta \op{a}^\dagger \op{a}}
    \end{equation}
generates an anti-clockwise rotation by $\theta$ in phase space. A measurement of $\op{p}_\theta$ with outcome $m$, realized via homodyne detection, corresponds to a projection onto the quadrature eigenstate,
\begin{equation} \label{scrivauto:43}
\begin{aligned}
\ketsub{m}{p_{\theta}} \coloneqq \op R^\dagger(\theta) \pket{m} \, ,
\end{aligned}
\end{equation}
where $ \op{p}_{\theta}  \ketsub{m}{p_{\theta}} = m \ketsub{m}{p_{\theta}}.$\footnote{The phase-delayed eigenstates are defined via a Heisenberg-picture transformation, just as in Refs.~\cite{Menicucci2006, Alexander2014, Alexander2018}.}

Measuring both local modes in a single macronode in rotated momentum quadratures given by measurement angles $\theta_a$ and $\theta_b$ teleports the state to the next macronode and implements the operation  \cite{Alexander2016, Yokoyama2013, Ukai2010}
 
    \begin{equation} \label{oldschoolKraus}
        \op{D}(\mu) \op{V}(\theta_a, \theta_b) \, ,       
    \end{equation}
with measurement-basis-dependent Gaussian unitary
    \begin{align} \label{Vgate}
        \op V(\theta_a, \theta_b) \coloneqq
         \op{R}(\theta_+ - \tfrac{\pi}{2}) \op{S}(\tan \theta_- ) \op{R}(\theta_+),
    \end{align}
and displacement $\op{D}(\mu)$ with outcome-dependent amplitude 
    \begin{equation} \label{mainmu}
        \mu =- \frac{ m_a e^{i \theta_b} + m_b e^{i \theta_a } }{ \sin (\theta_a - \theta_b)}\, ,
    \end{equation}
which can be corrected with active Gaussian shifts.\footnote{Various works have slightly different definitions of these operations. A derivation of Eq.~\eqref{oldschoolKraus} and further discussion can be found in Appendix \ref{appendix:Vop}.}
The parameters 
    \begin{equation} \label{measurementangles}
        \theta_\pm \coloneqq \tfrac{1}{2}( \theta_a \pm \theta_b)
    \end{equation}
are symmetric and antisymmetric combinations of the measurement angles, and we define a (nonstandard) squeezing operator,
    \begin{equation} \label{squeezingop}
        \op{S}(\zeta) \coloneqq \op{R}(\Im \ln \zeta) e^{-\frac{i}{2} (\Re \ln \zeta) (\op{q} \op{p} + \op{p}\op{q})}.
    \end{equation}
This is just an ordinary squeezing operator with squeezing parameter $r = \ln |\zeta|$ generalized to allow negative values of $\zeta$, which can arise in Eq.~\eqref{Vgate}. For $\zeta<0$, Eq.~\eqref{squeezingop} describes squeezing followed by a $\pi$ phase delay, which is equivalent to a double Fourier transform or a parity operation.\footnote{This squeezing operator is designed such that $\op{S}^\dagger(\zeta) \op{q} \op{S}(\zeta) = \zeta \op{q}$ and $\op{S}^\dagger(\zeta) \op{p} \op{S}(\zeta) = \zeta^{-1} \op{p}$ for all $\zeta \in \mathbb{R}_{\neq 0}$.} Note that this operator produces squeezing (and conjugate antisqueezing) along the principal position and momentum axes; squeezing other quadratures is achieved by following this operator by a phase delay, Eq.~\eqref{phasedelay}. Note that two teleportation steps through the macronode wire (Figure~\ref{macronodecartoon}) are sufficient to perform two different gates of the form of Eq.~\eqref{oldschoolKraus} and thus an arbitrary Gaussian unitary~\cite{Alexander2014}.

\subsection{Connection to maximally entangled EPR states}

In the ideal macronode CVCS construction, the Kraus state is composed of local modes, shared between macronodes, prepared in 0-momentum and 0-position eigenstates, Eq.~\eqref{scrivauto:771}, and then coupled on a beam splitter. Here, we show that the resulting state is a member of a complete set of maximally entangled two-mode states. 

Position and momentum eigenstates each form a resolution of the identity over a single mode, thus it is straightforward to construct a tensor-product representation of the identity across two modes. Consider the following tensor-product basis,
    \begin{align} \label{tensprodbasis}
        \op{I}_1 \otimes \op{I}_2  = \iint ds \, dt \, \outprodsubsub{t}{t}{p_1}{p_1}\otimes \outprodsubsub{s}{s}{q_2}{q_2}\, .
    \end{align}
Any unitary transformation on this expression also gives a resolution of the identity and produces a new, typically entangled, basis. Specifically, we consider the unitary $e^{-i\op{q} \otimes \op{p} }$, which serves as the CV analog to a CNOT gate \cite{Bartlett2002}, to define a complete set of shifted maximally entangled EPR states,
    \begin{subequations} \label{EPRalt}
    \begin{align}
        \ket{\EPR(s,t)} \coloneqq & e^{-i\op{q}_1 \otimes \op{p}_2 } \ket{t}_{p_1} \otimes \ket{s}_{q_2} \\
            = & \frac{1}{ \sqrt{2\pi} } \int dr \, e^{irt}  \ket{r}_{q_1} \otimes \ket{s + r}_{q_2}   \, ,
    \end{align}
    \end{subequations}
satisfying $\inprod{\EPR(s,t)}{\EPR(s',t')} = \delta(s-s') \delta(t-t')$.
The set of shifted EPR states resolves the identity over two modes,
    \begin{align}
    \label{eq:EPRcomplete}
        \op{I}_1 \otimes \op{I}_2  = \iint ds \, dt \, \outprod{\EPR(s,t)}{\EPR(s,t)},
    \end{align}
comprising an entangled two-mode basis that complements the tensor-product basis above, Eq.~\eqref{tensprodbasis}.

The states in Eq.~\eqref{EPRalt} can also be written as a momentum shift, $\op{Z}(t) \coloneqq e^{it\op{q}}$, and a position shift, $\op{X}(s) \coloneqq e^{-is \op{p}}$, across the two modes as
    \begin{equation} \label{EPRshiftpull}
        \ket{\EPR(s,t)} =  \op{Z}_1(t) \op{X}_2(s) \ket{\EPR}   \, ,
    \end{equation}
where $\ket{\EPR} \coloneqq \ket{\EPR(0,0)}$, is a \emph{canonical} EPR state, with position-position representation,
    \begin{equation}\label{EPR}
        \ket{\EPR} \coloneqq \frac{1}{\sqrt{2 \pi}} \int dr\, \ket{r}_{q_1} \otimes \ket{r}_{q_2} \, .
    \end{equation}
This state, being perfectly correlated in position and perfectly anti-correlated in momentum, is the continuous-variable analog of the Bell state $\ket{\Phi^+} = \frac{1}{\sqrt{2}} \sum_{j=0,1} \ket{j} \otimes \ket{j}$. The canonical EPR state can be represented as $\ket{\EPR} = \op{C}^X_{12}(1) \pket{0} \otimes \qket{0}$,
where the entangling gate is a CV controlled-$X$ gate of weight $g$,
    \begin{equation} \label{controlledX}
        \op{C}^X_{jk}(g) \coloneqq e^{  -i g \op{q}_j \otimes \op{p}_k }\, ,
    \end{equation}
with control mode $j$ and target mode $k$. This gives the circuit identity:
\begin{equation} \label{EPRCX}
       \begin{split}
 \Qcircuit @C=0.3cm @R=0.3cm {
         &&&&&&& \ctrl{1}  & \rstick{\pket{0}} \qw  &&&&                 &&&  \qw & \ar @{-} [dr(0.5)]\qw  &\\
         &&\ustick{\raisebox{.2em}{$\smash{\ket\EPR} \;\coloneqq$}}&&&&& \targ  & \rstick{\qket{0}}  \qw &&&& \ustick{\raisebox{.2em}{$=$}} &&& \gate{\frac{1}{\sqrt{2\pi}} I } & \ar @{-} [ur(0.5)] \qw  & 
    } 
       \end{split}\, .
\end{equation}
The factor $\frac{1}{\sqrt{2\pi}}$ results from the normalization of the EPR state in Eq.~\eqref{EPR}.\footnote{For qudits, the normalization of the maximally entangled states used for teleportation, $\frac{1}{\sqrt{d}}$, is typically absorbed into the ``angle-bracket'' notation for these states in a circuit diagram~\cite{Gottesman1999} and also ignored in the teleported gate. For CV systems considered here, we make the association $d \rightarrow 2\pi$ (for $\delta$-normed states) and explicitly include this EPR normalization in teleported gates so that the resulting Kraus operators are properly normalized.}
Note that we use the convention that quantum circuits proceed in time from right-to-left, such that a circuit-description of a state respects the same ordering as its partner equation.

\subsubsection{Bouncing operators on an EPR state}

Let $\op{O}$ be a single-mode operator with position-space representation
\begin{align}
    \hat O
&=
    \iint ds\, ds'\, O(s,s') \qoutprod{s}{s'}\,,
\end{align}
where $O(s,s') = \qbra s \op O \qket {s'}$. Applying this operator to one mode of an EPR state gives
    \begin{subequations} \label{eq:Choistate}
    \begin{align} 
        \ket{\Psi_O} & \coloneqq (\hat O \otimes \hat I) \ket{\EPR} \\
        & = \frac{1}{\sqrt{2 \pi}} \iint ds\, ds'
        \, O(s,s') \ketsub{s} {q_1} \otimes \ketsub {s'}{q_2} \, .
    \end{align}
    \end{subequations}
The state $\ket{\Psi_O}$ is called the \emph{Choi state} for $\op O$ and is a representation of $\op{O}$ through the Choi-Jamio\l{}kowski isomorphism~\cite{Jam1972,Choi1975}. It derives from the position-space expansion of~$\op O$ simply by replacing $\qoutprod s {s'} \to \ketsub s {q_1} \otimes \ketsub {s'} {q_2}$ (up to normalization). 
Teleporting through a Choi state applies the associated operator, which we call the teleported gate \cite{Gottesman1999}. This forms the basis for gate teleportation and the Kraus operator derived in Sec.~\ref{genmacronode}.

An important feature of maximally entangled states is that a local operator acting on one mode can be moved to the other (the operator is modified in the process), and the resulting two-mode state is identical to the original. 
This operation, which we call \emph{bouncing}, is described by the circuit, 
\begin{align}\label{scrivauto:88}
    \begin{split}
    \Qcircuit @C=1em @R=1em {
    &\qw & \gate{O} & \qw[-1] \ar @{-} [dr(0.5)] & & & &\qw & \qw & \qw[-1] \ar @{-} [dr(0.5)] &\\
    &\qw & \qw & \qw[-1] \ar @{-} [ur(0.5)] & & \raisebox{2em}{$=$} & &\qw & \gate{O^{\tp}} & \qw[-1] \ar @{-} [ur(0.5)] & \, \\
    }
    \end{split},
\end{align}
where $\op{O}$ is an operator on the top mode, and $\op{O}^\tp$ (the bounced operator) is the transpose of $\op{O}$ taken in the basis where the maximally entangled state is perfectly correlated---the position-position basis for the EPR state in Eq.~\eqref{EPR}.
Thus, the Choi state above can be represented in different, yet equivalent ways. 

The transposed position and momentum operators, $\op{q}^\tp = \op{q}$ and $\op{p}^\tp = -\op{p}$, can be found by computing their matrix elements in the position basis,
\begin{equation}\label{scrivauto:89}
    \begin{aligned}
      q^{\tp}_{(t,s)}  &= t \, \delta(t-s) = q_{(t,s)} \\
      p^{\tp}_{(t,s)}  &= i\, \delta'(t-s) = -p_{(t,s)} ,
    \end{aligned}
\end{equation}
where $x_{(t,s)} \coloneqq \qbra{t} \op{x} \qket{s} $.\footnote{The relations follow from the symmetry of $\delta(\cdot)$ and the asymmetry of its derivative $\delta'(\cdot)$.} 
These relations allow a straightforward bounce of a displacement operator, 
    \begin{equation}
        \op{D}(\alpha) \coloneqq e^{\alpha \op{a} - \alpha^* \op{a}^\dagger} = e^{i \sqrt{2} (\alpha_I \op{q} - \alpha_R \op{p})} \, , 
    \end{equation}
with complex phase-space displacement $\alpha = \alpha_R + i \alpha_I$, through the EPR state in Eq.~\eqref{EPR}:
	\begin{equation}
		\op{D}_1(\alpha) \ket{\EPR}
		=  \op{D}_2(-\alpha^*)  \ket{\EPR} \label{dispbounce} .
	\end{equation}
More general single-mode operators generated by powers of $\op{q}$ and $\op{p}$ can also be bounced using Eqs.~\eqref{scrivauto:89}. 
Note that Eq.~\eqref{dispbounce} can straightforwardly be used to bounce position and momentum shifts by decomposing the displacement operator,
    \begin{equation} \label{Ddecompose}
        \op{D} (\alpha) = e^{i \alpha_R \alpha_I} \op{X}\big(\sqrt{2} \alpha_R \big) \op{Z}\big(\sqrt{2} \alpha_I \big) \, .
    \end{equation}

\subsubsection{Entangled states on a beam splitter}

In a macronode wire, pairs of modes---one each from neighboring macronodes---are combined on a beam splitter. When these modes are prepared in position and momentum eigenstates, respectively, the result is a shifted EPR state, Eq.~\eqref{EPRalt}, whose entanglement lies at the heart of continuous-variable teleportation protocols. 

Consider momentum and position eigenstates, $\ket{t}_{p}$ and $\ket{s}_{q}$, coupled by a beam splitter. As shown in Appendix \ref{appendix:beamsplitterstates}, the resulting entangled state is
\begin{equation}
    \bsop_{12} \ket{t}_{p_1} \otimes \ket{s}_{q_2}
     =\sqrt{2} e^{ist} \ket{\EPR(\sqrt{2}s,\sqrt{2}t)}\, , \label{BSalmostthere} 
\end{equation}
up to a fixed phase (which is irrelevant). The norm scaling by~$\sqrt{2}$ preserves inner products and ensures that the completeness relation, Eq.~\eqref{eq:EPRcomplete}, holds,
\begin{equation}
      \op{I}_1 \otimes \op{I}_2 
=
    \iint dt \, ds \, \big( \bsop_{12} \ketsub t {p_1} \otimes \ketsub s {q_2} \big) \big(\brasub t {p_1} \otimes \brasub s {q_2} \bsop_{12}^\dag \big).
\end{equation}
We keep track of this factor throughout so that the Kraus operators we derive in Sec.~\ref{genmacronode} can be used to faithfully calculate measurement probabilities.

A final manipulation of the state in Eq.~\eqref{BSalmostthere} provides a useful form for the beam splitter entangled state.
Pulling out the shifts [Eq.~\eqref{EPRshiftpull}], bouncing the momentum shift $\op{Z}_1(\sqrt{2}t)$ to the second mode [Eq.~\eqref{dispbounce}], and combining the two single-mode shifts into a displacement operator [Eq.~\eqref{Ddecompose}] gives
	\begin{align}	 \label{beamsplitterquadstates}
		\bsop_{12} \ket{t}_{p_1} \otimes \ket{s}_{q_2} 
			= &\sqrt{2} \op{D}_2 ( s+it ) \ket{\EPR} \, .
	\end{align}
This state is represented by the circuit diagram,
  \begin{equation}\label{entangleBS}
        \begin{split}
		 \Qcircuit @C=1em @R=1.5em {
         	& \bsbal{1} & \rstick{\pket{t}} \qw &&&&                     &&& \qw                                                               & \ar @{-} [dr(0.5)]\qw  &\\
         	& \qw       & \rstick{\qket{s}} \qw &&&& \raisebox{2em}{$=$} &&& \gate{\frac{1}{\sqrt{\pi}} D(s+it) } & \ar @{-} [ur(0.5)] \qw &
  		  } \, 
  		\end{split} \, ,
	\end{equation}
indicating that the beam splitter entangled state is the Choi state, Eq.~\eqref{eq:Choistate}, for the displacement $\op{D}(s+it)$ (up to a factor $\tfrac{1}{\sqrt{\pi}}$).

Thus, the specific entangled state between adjacent macronodes in Eq.~\eqref{scrivauto:771},
$\bsop_{12} \ket{0}_{p_1} \otimes \ket{0}_{q_2} = \sqrt{2} \ket{\EPR}$,
is the Choi state \cite{Choi1975} for the identity, which underpins its utility for teleportation from one macronode to the next in the standard macronode computing protocol.

\section{Gate teleportation with the macronode wire \label{genmacronode}}

The macronode procedure for quantum computation, described in Sec.~\ref{sec:MBQCmacronode}, has two fundamental components. The first is the multimode entangled state itself---the macronode wire---and the second is the collection of homodyne measurements. Both of these components are Gaussian, which limits effective operations on an input state to Gaussian operations. Here, we show that modifying one part of this procedure, the state of the macronode wire itself, allows us to implement non-Gaussian operations in a gate-teleportation fashion. The generalized macronode procedure is shown in Figure \ref{macronodecartoon}. The key difference from the standard procedure is the replacement of the quadrature eigenstates that comprise a CVCS macronode wire [see Eq.~\eqref{scrivauto:771}] with arbitary pure states. Modes from adjacent macronodes are still measured in rotated homodyne bases, with the ultimate effect that the Gaussian operation, Eq.~\eqref{Vgate}, is supplemented with an additional ancilla-state-dependent operation. This additional operation can be leveraged for many purposes, including teleportation-based error correction on a CV-encoded qubit (using the Gottesman-Kitaev-Preskill code), which we describe in Sec.~\ref{GKP}.

\begin{figure}[t]
\centering
\resizebox{.8\columnwidth}{!}
   {
        \Qcircuit @C=0.35cm @R=0.75cm 
        {
	&&&&&&&&&	&&&& &&& \lstick{\cdots}   &  \bsbal{1}  &  \rstick{ \text{(in)}} \qw & \\
	&&&&&&&&	& \lstick{\brasub{m_{a}}{p_{\theta_a}}} & \qw & \bsbal{1} & \qw       & \qw & \qw & \qw & \qw & \qw    & \rstick{ \cdots } \qw & \\
	&&&&&&&&	& \lstick{\brasub{m_{b}}{p_{\theta_b}}}& \qw  & \qw       & \bsbal{1} & \rstick{ \ketsub{\psi}{} } \qw &&&&&& \\
		 & \lstick{\cdots} & \bsbal{1} &\qw&\qw & \qw &\qw &\qw &\qw &\qw &\qw  &\qw &  \qw  &  \rstick{ \ketsub{\phi}{}} \qw &&&&&& \\
		 & \lstick{\text{(out)}} & \qw  & \rstick{ \cdots } \qw \gategroup{2}{5}{4}{16}{2em}{--} & & & &&&&&&&&&&&& \\
		}
	}
\caption{
Circuit diagram for gate teleportation using the macronode wire in Fig.~\ref{macronodecartoon}. 
In our convention, circuits proceed from right to left.
 Inside the dashed box is the teleportation gadget, which consists of three modes. Arbitrary ancillae on the second and third mode, $\ket{\psi}$ and $\ket{\phi}$, are combined on a beam splitter to form the \emph{Kraus state}. The first two modes, which comprise a single macronode, are combined on an additional beam splitter and measured via homodyne detection in rotated bases. We describe these measurements as projections onto rotated momentum eigenstates, Eq.~\eqref{scrivauto:43}. 
 This procedure teleports a state from the first mode to the third mode with a (potentially nonunitary) gate applied that depends on the measurement bases, the measurement outcomes, and the Kraus state.
}\label{macronodecircuit}
\end{figure}
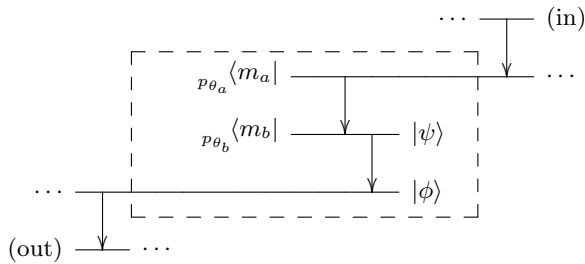

\subsection{Kraus operator for gate  teleportation}

We focus on the teleportation gadget within the macronode wire, indicated in the dashed box in Fig.~\ref{macronodecircuit}. This gadget describes the teleportation of a quantum state arriving from the previous macronode through the given macronode using two homodyne measurements after a beam splitter---an effective EPR measurement. This teleportation is accompanied by the Gaussian operation $\op V(\theta_a, \theta_b)$ in Eq.~\eqref{Vgate}, determined by the bases in which the homodyne measurements are performed. As we show here, a non-Gaussian and non-unitary operation can also be applied by preparing local modes of adjacent macronodes in the arbitrary pure, product state $\ket{\psi} \otimes \ket{\phi}$ and then mixing them on a beam splitter. The entire operation is described by the Kraus operator,
    \begin{equation} \label{krauseq}
        \op{K}(m_a,m_b) \coloneqq \brasub{m_a}{p_{1,\theta_a}} \otimes \brasub{m_b}{p_{2,\theta_b}} \bsop_{12} \bsop_{23} \ket{\psi}_2 \otimes \ket{\phi}_3 \, ,
    \end{equation}
where $m_a$ and $m_b$ are the outcomes of the homodyne measurements, and subscripts in each beam splitter, $\bsop_{jk}$, label the modes it couples. 
This Kraus operator is described by the circuit
    \begin{align}\label{scrivauto:77}
    \begin{split}
    \Qcircuit @C=1.75em @R=2em 
    {
	&\lstick{\brasub{m_a}{p_{\theta_a}}}   & \bsbal{1} & \qw & \rstick{\text{(in)}} \qw[-1] &  \\
	&\lstick{\brasub{m_b}{p_{\theta_b}}}  & \qw   &  \bsbal{1} & \rstick{\ket{\psi}} \qw \\
	&\lstick{\text{(out)}}	& \qw  &   \qw  & \rstick{ \ket{\phi} } \qw \\
		}\, 
	\end{split}	
    \end{align}
Again, we use the convention that time proceeds right-to-left in circuit diagrams.
We refer to the joint state of the last two modes, $\bsop_{23} \ket{\psi}_2 \otimes \ket{\phi}_3$, as the \emph{Kraus state}, which is the Choi state for a (typically non-unitary) teleported gate whose form we determine below.

After an input state on the top wire $\op{\rho}_\text{in}$ traverses the teleportation gadget as part of a macronode wire, it is teleported to the third mode and transformed according to the conditional map,
    \begin{equation} \label{Krausmapdef}
        \op{\rho}_\text{out} = \frac{\op{K}(m_a,m_b) \op{\rho}_\text{in} \op{K}^\dagger(m_a,m_b)}{\Pr(m_a,m_b)}\, ,
    \end{equation} 
where the joint probability (density) of obtaining the outcomes $\{ m_a, m_b \}$ is
    \begin{equation} \label{jointprob}
        \Pr(m_a,m_b) = \Tr [\op{K}^\dagger(m_a,m_b)\op{K}(m_a,m_b) \op{\rho}_\text{in}] \, .
    \end{equation} 
As the modes in the macronode wire are successively measured, a sequence of such Kraus operators is applied to the input state, each with a set of measurement outcomes and an applied operation depending on which quadratures are measured and the ancillae at the beam splitter, which comprise the Kraus state. It suffices to consider the single Kraus operator in Eq.~\eqref{krauseq}, as it contains all the necessary ingredients.

The homodyne measurements on the first two modes realize an entangled Bell-type measurement with the extension that, by measuring rotated quadratures, the basis-dependent Gaussian operation $\op{V}(\theta_a, \theta_b)$ in Eq.~\eqref{Vgate} is applied. This is described by the following circuit identity 
\begin{equation}\label{measurementcircuit}
\begin{split}
\centering
\resizebox{.8\columnwidth}{!}{
    \Qcircuit @C=1em @R=1.5em 
    {
	&&\lstick{\brasub{m_a}{p_{\theta_a}}} &\bsbal{1} &\qw &                      &&\ar @{-} [dl(0.5)] &\gate{D(\mu)} &\gate{\frac{1}{\sqrt{\pi}}V(\theta_a ,\theta_b)} & \qw \\
	&&\lstick{\brasub{m_b}{p_{\theta_b}}} &\qw       &\qw &\raisebox{2.5em}{$=$} &&\ar @{-} [ul(0.5)] &\qw           &\qw                                              & \qw \\
	}
	}
	\end{split}
\end{equation}
where the effect of the outcomes is a displacement by $\mu$, given in Eq.~\eqref{mainmu}.

We now focus on the Kraus state on the second and third modes $\bsop_{23} \ket{\psi}_2 \otimes \ket{\phi}_3$. First, we write the state $\ket{\psi} \otimes \ket{\phi}$ in the tensor-product basis, Eq.~\eqref{tensprodbasis}, using the respective momentum and position wavefunctions,
    \begin{equation}
        \label{eq:wavefunctions}
        \tilde\psi(t) \coloneqq \inprodsubsub{t}{\psi}{p}{} \qquad \text{and} \qquad \phi(s) \coloneqq \inprodsubsub{s}{\phi}{q}{} \, .
\end{equation}
Then, we apply the beam splitter using Eq.~\eqref{beamsplitterquadstates} to arrive at an expression for the Kraus state in terms of an operation on an $\ket{\EPR}$ state,
	\begin{align}	
			\bsop_{23} \ket{\psi}_2\otimes \ket{\phi}_3 & = \bsop_{23} \iint dt ds \, \tilde{\psi}(t) \phi(s) \ket{t}_{p_2} \otimes \ket{s}_{q_3} \\
			& =  \sqrt{2} \bounceEPR{\psi, \phi}{3}  \ket{\EPR} \, .
	\end{align}	
The single-mode operator $\op{A}(\psi,\phi)$, referred to as the \emph{teleported gate}, arises in its Cahill-Glauber form~\cite{Cahill1969,Cahill1969a}, with displacements weighted by the respective momentum and position wavefunctions of the input states,
	\begin{equation} \label{EPRbounce}
		\bounceEPR{\psi, \phi}{} \coloneqq  \iint  d^2 \alpha \,  \tilde{\psi} ( \alpha_I ) \phi ( \alpha_R ) \op{D}  (\alpha ),
	\end{equation}
for complex number $\alpha = \alpha_R + i \alpha_I$. 
The Kraus state can be reexpressed with the circuit identity, 
\begin{equation} \label{ancillacircuit}
\begin{split}
 \Qcircuit @C=0.5cm @R=0.5cm 
    {
         & \bsbal{1} & \rstick{\ket{\psi}} \qw &&                               && \qw                                                       & \ar @{-} [dr(0.5)] \qw &\\
         & \qw       & \rstick{\ket{\phi}} \qw && \ustick{\raisebox{.5em}{$=$}} && \gate{ \frac{1}{\sqrt{\pi}} \bounceEPRgate{\psi, \phi}{}} & \ar @{-} [ur(0.5)] \qw & 
    } 
\end{split} \, ,
\end{equation}
where the $\frac{1}{\sqrt{\pi}}$ arises due to the circuit identity for the $\ket{\EPR}$ state, Eq.~\eqref{scrivauto:771}. In this form, it is clear that the operator $\bounceEPR{\psi, \phi}{}$ is indeed a teleported gate in the context of the standard teleportation gadget \cite{Gottesman1999}.

Combining the circuits for the measurements, \eqref{measurementcircuit}, and the Kraus state, \eqref{ancillacircuit}, we express the teleportation gadget in Eq.~\eqref{scrivauto:77} as
\begin{align}\label{EPRstate}
\begin{split}
    \Qcircuit @C=1em @R=1em {
    &\ar @{-} [dl(0.5)] &\gate{ D(\mu) } &\gate{\frac{1}{\sqrt{\pi}} V(\theta_a ,\theta_b)}      &\rstick{\text{(in)}} \qw[-1] &     \\
    &\ar @{-} [ul(0.5)] &\qw             &\qw                                                    &\qw[-1] \ar @{-} [dr(0.5)] &&   \\
	&\lstick{\text{(out)}}       &\qw             &\gate{\frac{1}{\sqrt{\pi}}\bounceEPRgate{\psi,\phi}{}} &\qw[-1] \ar @{-} [ur(0.5)] &&   \\
		}
\end{split}		
\end{align}
Pulling the circuit taut,
\begin{equation}\label{scrivauto:81}
\begin{split}
\resizebox{.71\columnwidth}{!}{
    \Qcircuit @C=0.75em @R=1em {
        &\lstick{\text{(out)}} &\gate{\frac{1}{\sqrt{\pi}}\bounceEPRgate{\psi,\phi}{}} &\gate{ D(\mu) } &\gate{\frac{1}{\sqrt{\pi}}V(\theta_a ,\theta_b)} &\rstick{\text{(in)}}\qw \\
	} 
	}
\end{split}
\end{equation}
allows us to read off the Kraus operator directly by virtue of the right-to-left circuit convention,
    \begin{equation} \label{genkraus}
        \op{K}(m_a,m_b) = \frac{1}{\pi} \bounceEPR{\psi, \phi}{} \op{D}(\mu) \op{V}(\theta_a,\theta_b) .
    \end{equation}
While the first two operations, $\op{D}(\mu)$ and $\op{V}(\theta_a, \theta_b)$ [Eq.~\eqref{Vgate}], are Gaussian, the teleported gate $\bounceEPR{\psi, \phi}{}$ [Eq.~\eqref{EPRbounce}] can realize non-Gaussian operations when the ancilla states, $\ket{\psi}$ and $\ket{\phi}$, are themselves non-Gaussian. In addition, the operation realized by the teleported gate can be non-unitary. We apply this powerful tool below to realize teleportation-based error correction of a bosonically encoded qubit. 

Finally, we note that standard teleportation is achieved when the Kraus state is the canonical EPR state, and the measurement angles are chosen to be $\theta_a=\frac{\pi}{2}$ and $\theta_b=0$ so that $\op V(\theta_a,\theta_b)=\op I$.

\subsubsection{Damped-ancillae Kraus operator} \label{sec:dampedancillae}

The Kraus operator, Eq.~\eqref{genkraus}, implements operations determined by the wavefunctions of the input ancillae that comprise the Kraus state. 
In macronode CVCS quantum computing, each ancilla is prepared in a squeezed vacuum state \cite{Menicucci2006} that approximates a quadrature eigenstate to a degree that depends on the level of squeezing.
In a standard description, these states are described by unitarily squeezing the vacuum state, via $\op{S}(\zeta)$ in Eq.~\eqref{squeezingop}. This gives the squeezed vacuum state, which we write in the notation of Ref.~\cite{Menicucci2006},
    \begin{equation} \label{qsqueezedstate}
        \qket{0; \zeta} \coloneqq \frac{1}{ (\pi \zeta^2)^{1/4} }\int ds \, e^{\frac{-s^2}{2 \zeta^2}} \qket{s}
         = \op{S}(\zeta) \ket{0}\,,
    \end{equation}
where $\ket 0$ (with no subscript) is the vacuum state, and $\op{S}(\zeta)$ is the squeezing operator in Eq.~\eqref{squeezingop}, with squeezing factor~$\zeta$. The measured variance of this state is~$\avg{\op q^2} = \zeta^2/2$. $\zeta = 1$ reproduces the vacuum state, and the limit~$\zeta \to 0^+$ gives a normalized version of~$\qket 0$. We also define
    \begin{equation} \label{squeezedstate}
        \pket{0; \zeta} \coloneqq \frac{1}{ (\pi \zeta^2)^{1/4} }\int ds \, e^{\frac{-s^2}{2 \zeta^2}} \pket{s}
         = \op{S}(\zeta^{-1}) \ket{0}\,,
    \end{equation}
which analogously has $\avg{\op{p}^2} = \zeta^2/2$. Note that this implies that \begin{align}
    \pket{0; \zeta} = \qket{0; \zeta^{-1}}\,.
\end{align}

It will be helpful to list a few important relations for use in what follows. First, let
\begin{align}
    \zeta = e^{-r}\,.
\end{align}
Then, in the Fock basis~$\{ \ket n \}_{n=0}^\infty$,
\begin{subequations}
\label{eq:qsqueezewithr}
\begin{align}
    \qket{0; \zeta}
&=
    \sqrt{\sech r} \sum_{n=0}^\infty
    (-\tanh r)^n \frac{\sqrt{(2n)!}}{2^n n!}
    \ket{2n}
\\*
&=
    \sqrt{
    \frac{2\zeta}{1+\zeta^2}
    }
     \sum_{n=0}^\infty
    (-1)^n
    \left(\frac{1-\zeta ^2}{1+\zeta ^2}\right)^n \frac{\sqrt{(2n)!}}{2^n n!}
    \ket{2n}
    \,.
\end{align}
\end{subequations}
Also, $\pket {0; \zeta}$ is the same but
without the $(-1)^n$. Finally, by recognizing that $\inprodsubsub n 0 {} q = \psi_n^*(0)$, where $\psi_n$ is the $n$th Fock-state wave function, we can write the delta-normalised $\op q$-quadrature eigenstate in the Fock basis:
\begin{align}
\label{eq:qsqueezedelta}
    \qket 0
&=
    \pi^{-1/4} \sum_{n=0}^\infty
    (- 1)^n \frac{\sqrt{(2n)!}}{2^n n!}
    \ket{2n}\,,
\end{align}
and similarly for $\pket 0$ but again without the $(-1)^n$. These separate expressions are required because the prefactor vanishes in the normalised case, Eqs.~\eqref{eq:qsqueezewithr}, when ${\zeta \to 0^+}$ (${r \to \infty}$).

We now introduce an equivalent, non-unitary formulation of squeezed vacuum states that ultimately yields a convenient form for the associated macronode Kraus operator. Moreover, this method gives a straightforward way to include finite-squeezing noise in any state. Particularly useful cases are unphysical ideal states, such as the GKP states introduced in Sec.~\eqref{GKP}, and, more generally, any state constructed from a countable superposition of quadrature eigenstates.

The squeezed vacuum state in Eq.~\eqref{squeezedstate} can equivalently be described in terms of the non-unitary \emph{damping operator} defined by\footnote{The damping operator can also be considered a phase delay, Eq.~\eqref{phasedelay}, with imaginary delay angle $\theta = i \beta$.}
    \begin{equation} \label{damping}
        \dampop{\beta} \coloneqq e^{- \beta \op{n}}\,
    \end{equation}
acting on an infinitely squeezed quadrature eigenstate. To derive the precise relation, compare Eq.~\eqref{eq:qsqueezewithr} with Eq.~\eqref{eq:qsqueezedelta}. The only difference, aside from the normalisation, is a factor of $(\tanh r)^n$ in the sum. This factor can be restored by writing $(\tanh r)^{\op n/2} \qket 0$. Referring to Eq.~\eqref{damping}, we see that setting $\tanh r = e^{-2\beta}$ would produce the full relation
\begin{align}
\label{sqzstate}
        \qket{0; \zeta} &= \frac{1}{\sqrt{ \mathcal{N}_\zeta} } \dampop{\beta} \qket{0}
        \,,
\end{align}
with
\begin{align}
    \beta
&=
    \frac 1 2 \ln \coth r
=
    \frac 1 2 \ln
    \left(
        \frac{1+\zeta^2}{1-\zeta^2}
    \right)
    \,.
\end{align}
(Recall that $\zeta = e^{-r}$, as above.) The state is normalized by setting
    \begin{align} \label{sqznorm}
        \mathcal{N}_\zeta
    &=
        \frac{1}{\sqrt{\pi (1- e^{-4\beta})}}
    =
        \frac{\cosh r}{\sqrt\pi}
    =
        \frac{1+\zeta ^2}{2 \zeta \sqrt{\pi } }
        \,,
    \end{align}
which is required since $\dampop{\beta}$ is not unitary. Analogously, $\pket{0; \zeta} = \frac{1}{\sqrt{ \mathcal{N}_\zeta} } \dampop{\beta} \pket{0}$, with the exact same relations between $\zeta$, $r$, and $\beta$ as shown above.

 The damping operator has been used in macronode CVCS constructions previously, because it allows one to separate finite-squeezing effects from the ideal quadrature eigenstates \cite{Nick2014,alexander2016flexible,Alexander2017}. The damping operator can then be manipulated separately from the state on which it acts.
 This fact plays a critical role when the ancillae that comprise the Kraus state in the teleportation circuit in Eq.~\eqref{scrivauto:77} are described as damping operators acting on a state,
     \begin{align} \label{dampedstate}
        \ket{\psi} \rightarrow \frac{1}{ \sqrt{ \mathcal{N}_\psi}}\dampop{\beta} \ket{\psi}
    \end{align}
with normalization explicitly given by,
    \begin{align} \label{Krausnorm}
        \mathcal{N}_\psi \coloneqq \bra{\psi} \dampop{2\beta} \ket{\psi} .
    \end{align}
For identical damping on both ancillae, the joint damping operator  $\dampop{\beta} \otimes \dampop{\beta} = e^{-\beta(\op n_1 + \op n_2)}$ is a function of total photon number, $\op n_1 + \op n_2$, which is conserved by a beam splitter. The joint damping operator commutes trivially through the beam splitter in Eq.~\eqref{dampedancillacircuit}, and single-mode damping operators can then be bounced trivially to the final mode, since $\op{n}^\tp = \op{n}$ follows from Eq.~\eqref{scrivauto:89}. The resulting circuit identity is
	\begin{equation} \label{dampedancillacircuit}
\resizebox{1\columnwidth}{!}{
 \Qcircuit @C=0.25cm @R=0.3cm 
    {
         &\qw &\bsbal{1} & \qw &\gate{\dampgate} &\rstick{\ket{\psi}} \qw &&&             &&\qw              &\qw                                                      & \qw              & \ar @{-} [dr(0.5)] \qw  &\\
         &\qw &\qw       & \qw & \gate{\dampgate} &\rstick{\ket{\phi}} \qw &&& \ustick{\raisebox{.5em}{$=$}} &&\gate{ \dampgate} &\gate{ \frac{1}{\sqrt{\pi}} \bounceEPRgate{\psi, \phi}{}}& \gate{\dampgate} & \ar @{-} [ur(0.5)] \qw  & 
    }
    }
\end{equation}
Using this result in the macronode gadget, Eq.~\eqref{scrivauto:77}, gives the final Kraus operator
    \begin{align} \label{genkrausdamp}
        \op{K}&(m_a,m_b) = \nonumber \\
        &\frac{1}{\pi \sqrt{ \mathcal{N}_\phi \mathcal{N}_\psi }} \dampop{\beta} \bounceEPR{\psi, \phi}{} \dampop{\beta} \op{D}(\mu) \op{V}(\theta_a,\theta_b) \, ,
    \end{align}
with outcome-dependent displacement amplitude $\mu$, Eq.~\eqref{mainmu}. We emphasize here that this Kraus operator is a specific case of the general one derived above,  Eq.~\eqref{genkraus}, useful for situations when the ancillae states (that comprise the Kraus state) can be expressed as in Eq.~\eqref{dampedstate}.

\section{Application: GKP error correction in the macronode wire}

In its standard implementation, macronode-based quantum computing is used to perform Gaussian operations on an input state as it is teleported down the macronode wire \cite{Alexander2014}. Momentum and position quadrature eigenstates at the ancillae realize the intended Gaussian operation: $\op{D}(\mu) \op{V}(\theta_a ,\theta_b).$ Such states contain infinite energy and are not normalizable; in physical settings, finitely squeezed approximations to them are used. Pure squeezed states at the ancillae, described by Eq.~\eqref{sqzstate}, yield the Kraus operator
    \begin{equation} \label{genkrausdamp1}
        \op{K}(m_a,m_b) = \frac{1}{\pi  \mathcal{N}_{\zeta} } \dampop{2\beta} \op{D}(\mu) \op{V}(\theta_a,\theta_b) .
    \end{equation}
Finite-squeezing noise, originating in the ancillae and generated by $\dampop{2\beta}$, accompanies the Gaussian operation at each macronode. From a wavefunction perspective, this noise describes an equal blurring in position and momentum \cite{Nick2014,Alexander2014}. Given a sufficient number of teleportation steps, this noise overwhelms the process entirely, and any trace of the input state is wiped out. Further, a strict no-go theorem establishes that no procedure based on Gaussian operations can be used to correct this noise \cite{Niset2009}.   

The finite-squeezing noise that builds up through macronode teleportation (as well as other types of noise) can be dealt with using \emph{bosonic codes} \cite{Chuang1997,Cochrane1999,Wasilewski2007, Albert2018,Michael2016, Grimsmo2020}, which encode digital quantum information in the continuous Hilbert space of a mode. 
We focus on the Gottesman-Kitaev-Preskill (GKP) encoding of qubit into a mode \cite{GKP}, whose states and structure are given below. Important is the direct compatibility of the GKP code with the macronode wire (as well as with more sophisticated macronode protocols \cite{alexander2016flexible}), due to the fact that encoded Clifford gates and Pauli measurements are respectively realized by Gaussian unitaries and homodyne detection. Further, Gaussian operations suffice for universality and fault tolerance \cite{Baragiola2019, Yamasaki2020}. We show here that using GKP states as ancillae at various locations in a macronode wire performs GKP error correction, which projects built-up CV errors into potential logical errors.

\subsection{The GKP encoding}
\label{GKP}

For use in defining the GKP encoding, we first define a Dirac comb with spacing between spikes (period) of $T$ as
	\begin{equation}
		\Sha_{ T }(s) \coloneqq \sqrt{T} \sum_{n=-\infty}^\infty \delta(s - n T)\, .
	\end{equation}
The normalization of a Dirac comb is convenient because its Fourier transform takes the form, 
	\begin{equation}
		\mathcal{F} [\Sha_{ T }](t)  =  \sqrt{ \frac{2\pi}{T}} \sum_{n=-\infty}^\infty \delta \big( t - n \tfrac{2 \pi}{T} \big) = \Sha_{ 2\pi/T }(t) \, ,
	\end{equation}
with the Fourier transform defined as
    \begin{equation} \label{Fourier}
        \mathcal{F}[f](t) \coloneqq \frac{1}{\sqrt{2\pi}} \int_{-\infty}^\infty ds \, e^{-ist} f(s).
    \end{equation}
Scaling the argument of a Dirac comb by a real number $a$ gives a relation between Dirac combs of different period,
	\begin{equation} \label{shashifter}
		\Sha_{ T }( a s) 
		    = \tfrac{1}{\sqrt{a}} \Sha_{T/a}(s) \, ,
	\end{equation}
which follows using $\delta(as) = \frac{1}{a} \delta(s)$.

\begin{figure}[t!]
\centering
\includegraphics[width=1\columnwidth]{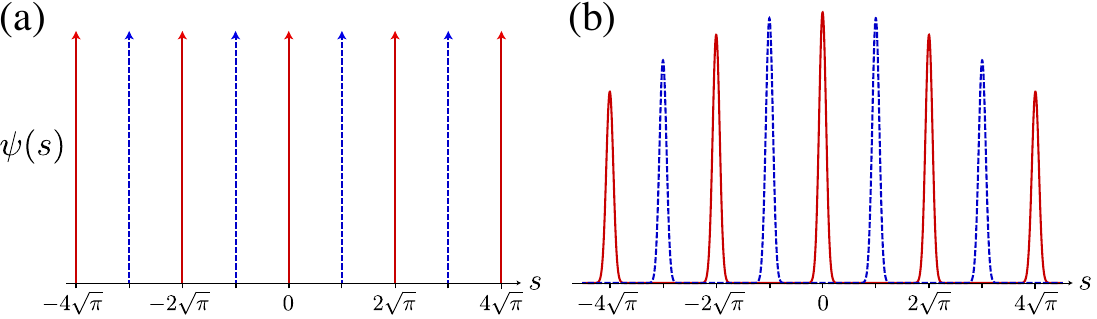}
\caption{Position wave functions for logical-0 (solid red) and logical-1 (dashed blue) GKP states. (a) Ideal GKP states. (b) Approximate GKP states with squeezing $s_\GKP$ = 18.6 dB. 
}\label{scrivauto:53}
\end{figure}

The ideal square-lattice GKP encoding for a qubit consists of two computational-basis states given by
	\begin{align} \label{GKPlogical}
		\ket{j_\GKP} &\coloneqq 
		\int ds\, \Sha_{2\sqrt\pi}( s-j\sqrt{\pi} ) \qket s
\\
&=
        (2 \sqrt \pi)^{1/2} \sum_{n=-\infty}^\infty \ketsub{(2n+j)\sqrt{\pi}}{q} \, , \label{GKPqbasis}
	\end{align}
with $j \in \{0,1\}$.
These states span a two-dimensional subspace, allowing a GKP qubit to be encoded as $\ket{\psi_\GKP} = c_0\ket{0_\GKP} + c_1 \ket{1_\GKP}$. The position-space wavefunctions for the states $\ket{j_L}$ are
	\begin{equation} \label{GKPposwf}
		\psi_{j,\GKP} (s) = \Sha_{ 2\sqrt{\pi} }( s-j\sqrt{\pi} )  
		,
	\end{equation}
using the definition above for~$\Sha_T$ (see Fig.~\ref{scrivauto:53}). This allows us to write the momentum-space wave functions for the computational basis states in Eq.~\eqref{GKPlogical} as
    \begin{equation}\label{momentumWFshah}
        \tilde{\psi}_{j,\GKP}(t) = \mathcal F[\psi_{j,\GKP}](t) = e^{ i j \sqrt{\pi} t} \Sha_{\sqrt{\pi}}(t),
    \end{equation}
which are Dirac combs with half the period of those in Eq.~\eqref{GKPposwf} and whose teeth have alternating phase for $\ket{1_L}$. Therefore, we have the momentum expansion
    \begin{align}
        \ket{j_\GKP} &=
    	\int dt\, e^{ i j \sqrt{\pi} t } \Sha_{\sqrt\pi}(t) \pket t
        \\
        &= (\sqrt \pi)^{1/2} \sum_{\mathclap{n=-\infty}}^\infty e^{i j \pi n } \pket{n\sqrt{\pi}} \label{GKPpbasis}
    \end{align}
From these expressions, position and momentum wavefunctions for the dual-basis logical states, $\ket{\pm_\GKP} \coloneqq \frac{1}{\sqrt{2}} \big( \ket{0_\GKP}\pm \ket{1_\GKP} \big)$, can be constructed. Using the wavefunctions above, $\inprod{j_\GKP}{k_\GKP} \propto \delta_{j,k}$, where the constant of proportionality is infinite. This is an artefact of using states with infinite norm. Physical approximations to these states will not have such pathologies.

The two-dimensional projector onto the GKP subspace is an important operator defined as
    \begin{equation} \label{GKPprojector}
        \GKPproj \coloneqq \outprod{0_\GKP}{0_\GKP} + \outprod{1_\GKP}{1_\GKP},
    \end{equation}
and can likewise be constructed from the dual-basis codewords $\ket{\pm_\GKP}$. 
Since the GKP eigenstates are a set of measure zero in the infinite-dimensional CV space, Eq.~\eqref{GKPprojector} is more appropriately a projector \emph{density} satisfying $(\GKPproj)^2 \propto \GKPproj$ with an infinite constant of proportionality. Nevertheless, in keeping with common usage, we still refer to $\GKPproj$ as a projector henceforth.\footnote{\label{foot:GKProj}In fact, similar behavior is seen for projectors onto position eigenstates: $(\qoutprod s s)^2 = \delta(0) \qoutprod s s$. These, too, are technically projector densities.}

A key feature of the square-lattice GKP code is that the Fourier transform operator $\op{F} = \op{R}(\tfrac{\pi}{2})$ (a phase delay, Eq.~\eqref{phasedelay}, by $\frac{\pi}{2}$) executes a logical Hadamard gate within the codespace. 
This is a result of the choice of spike period $T = 2\sqrt{\pi}$ for the position wavefunctions of the computational-basis states, Eq.~\eqref{GKPposwf}. 
By choosing a different period, $T=\sqrt{2\pi}$, we define another useful state,
    \begin{equation} \label{qunaught}
        \ket{\qunaught} := \int ds \; \Sha_{\sqrt{2\pi}}(s)\qket{s}
        =\int dt \; \Sha_{\sqrt{2\pi}}(t)\pket{t} \, .
    \end{equation}
The identical periodic structure in both position and momentum means that $\ket{\qunaught}$ is invariant under a Fourier transform, $\op{F}\ket{\qunaught} = \ket{\qunaught} $. This state was proposed in Duivenvoorden \emph{et al.} \cite{Duivenvoorden2017} as a sensor to simultaneously detect small shifts in position and momentum.
We focus on its quantum information properties and refer to $\ket{\qunaught}$ as a \emph{qunaught state} (in analogy to and with similar pronunciation as a "qubit") due to the fact that it defines a one-dimensional subspace and carries no quantum information.\footnote{One can, of course, define a rectangular GKP code for any spike spacing $T$ in the computational-basis codewords. In this case, executing a Hadamard gate typically requires both a Fourier transform and squeezing. 
Our focus here is the square-lattice code only.} 
Nevertheless, qunaught states serve as the resource for error correcting the square-lattice GKP code in the teleportation gadget, Eq.~\eqref{scrivauto:77}, as we show below.

\subsubsection{GKP Bell pairs}

\begin{figure}
 \centering
\includegraphics[width=1\columnwidth]{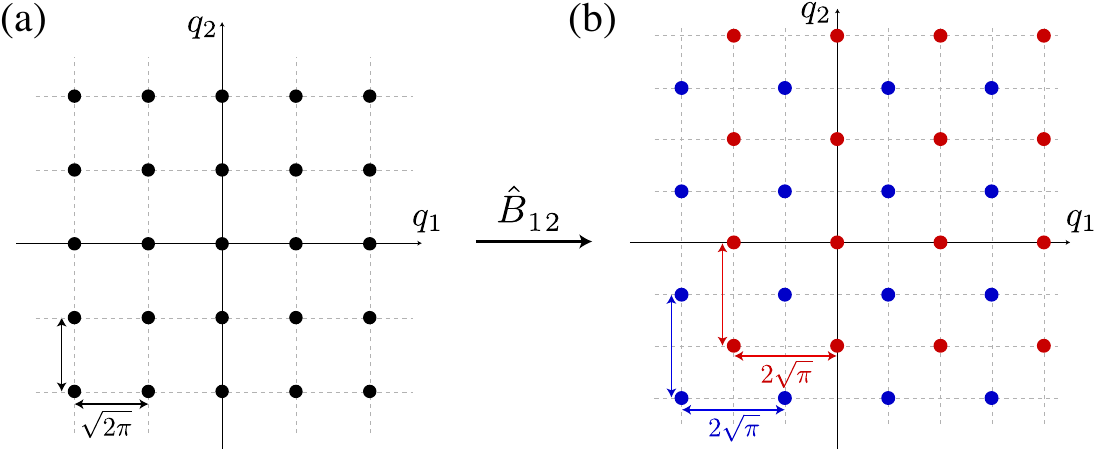}
 \caption{Generation of a GKP Bell pair by mixing two qunaught states, Eq.~\eqref{qunaught}, on a beam splitter. Shown in both figures is the two-mode position-position wavefunction, where each dot is a two-dimensional $\delta$-function. (a) Product state of two qunaughts, $\ket{\qunaught}\otimes\ket{\qunaught}$.  (b) Encoded GKP Bell state, Eq.~\eqref{GKPBellfromqunaught}, generated after the qunaughts are entangled by the beam splitter, which simply rotates the position-space wavefunction shown in~(a) by $\pi/4$ in the $(q_1,q_2)$-plane.
 This GKP Bell state is a sum of two product states, $\ket{0_\GKP}\otimes\ket{0_\GKP}$ and $\ket{1_\GKP}\otimes \ket{1_\GKP}$, indicated in red (solid) and blue (hollow), respectively.
}\label{gkpbell}
 \end{figure}

The GKP-encoded version of the qubit Bell state is 
\begin{equation}\label{gkpbellpair}
    \ket{\Phi^+_\GKP} \coloneqq \tfrac{1}{\sqrt{2}} \big( \ket{0_\GKP} \otimes \ket{0_\GKP} + \ket{1_\GKP} \otimes \ket{1_\GKP} \big).
\end{equation}
This state can be constructed on a beam splitter using two qunaught states, Eq. \eqref{qunaught}. First, we recognize that a pure state can be expressed as
    \begin{equation} 
    \label{eq:filterofq}
        \ket{\psi} = \int ds \; \psi(s)\qket{s} = \psi(\op{q}) \int ds \; \qket{s} = \sqrt{2\pi} \psi( \op{q} ) \ket{0}_p \, ,
    \end{equation}
 where $\psi(s)$ is the position wavefunction and $\int ds \qket{s} = \sqrt{2\pi} \pket{0}$. 
 Using this relation and the beam splitter transformation on the position operators, $\hat q_{\pm} = \tfrac{1}{\sqrt{2}} (\hat q_{1} \pm \hat q_{2})$ in Eq.~\eqref{bstransformation}, allows us to express the state as
\begin{align}
    &\bsop_{12} \ket{\qunaught} \otimes \ket{\qunaught}\nonumber \nonumber \\*
    &\quad = 2\pi \Sha_{\sqrt{2\pi}}(\op q_+) \Sha_{\sqrt{2\pi}}(\op q_-)         \pket{0}\otimes\pket{0}
    \nonumber \\
    &\quad = \frac{2 \pi}{\sqrt 2} \sum_{j=0}^1
        \Sha_{2\sqrt{\pi}}(\op q_1+j \sqrt{\pi})\Sha_{2\sqrt{\pi}}(\op q_2+j \sqrt{\pi})
        \pket{0}\otimes\pket{0} \nonumber \\*
    &\quad = \ket{\Phi^+_\GKP}\, . \label{GKPBellfromqunaught}
\end{align}
The generation of this state is shown graphically in Fig.~\ref{gkpbell}.
The beam splitter rotates the two-mode state in the $(q_1, q_2)$-plane by $\pi/4$, and the result is described by a sum of two separate lattices, each with Dirac-comb-period $2 \sqrt\pi$. 
The direct connection to a GKP Bell pair is found using Eq.~\eqref{GKPBellfromqunaught}.\blk

\subsubsection{Approximate GKP states}

Ideal GKP states, including the computational basis states in Eq.~\eqref{GKPlogical}, are unphysical, unnormalizable states. Physical approximations to these states replace each spike in the position wavefunction [Eq.~\eqref{GKPposwf}] with a sharp Gaussian, and then damp spikes far from the origin with a broad Gaussian envelope.\footnote{There are many other ways to approximate ideal GKP states, see Ref.~\cite{matsuura2019, Lucas2020} for more details.} This can be modeled mathematically by applying the damping operator, Eq.~\eqref{damping}, to the ideal states, which has the added benefit of treating the momentum wavefunction in an egalitarian way. 
The resulting approximate GKP states are
    \begin{equation}
        \ket{\bar{j}_\GKP} = \frac{1}{ \sqrt{ \mathcal{N}_{j,\GKP}} } \dampop{\beta} \ket{j_\GKP}\,.
    \end{equation}
The damping spoils the strict orthogonality of the codewords, $\inprod{\bar{0}_\GKP}{\bar{1}_\GKP} \neq 0$, which is a generic feature of various other physical bosonic codes, such as cat codes \cite{Lund2008}. Applying the damping operator to the GKP codespace projector, Eq.~\eqref{GKPprojector}, gives the (unnormalized) quasi-projector onto the subspace spanned by the approximate codewords,
    \begin{align} \label{dampGKPprojector}
        \approxGKPproj \coloneqq & \dampop{\beta} \big(\outprod{0_\GKP}{0_\GKP} +         \outprod{1_\GKP}{1_\GKP} \big) \dampop{\beta}\nonumber \\
        = & \mathcal{N}_{0,\GKP} \outprod{\bar{0}_\GKP}{\bar{0}_\GKP} + \mathcal{N}_{1,\GKP}\outprod{\bar{1}_\GKP}{\bar{1}_\GKP} \, .
    \end{align}
In the limit of ${\beta \rightarrow 0}$, this becomes the ideal GKP subspace projector.\footnote{For ${\beta \neq 0}$, the operator $\approxGKPproj$ is not a projector because $(\hat{\overline{\Pi}}{}_\GKP)^2 \not\propto \hat{\overline{\Pi}}{}_\GKP$. This is
due to the non-orthogonality of the approximate GKP codewords and is a separate issue from the infinite constant noted in the discussion below Eq.~\eqref{GKPprojector}.}

For small damping $\beta \ll 1$, the normalized position-space wavefunction for these states is~\cite{Lucas2020,matsuura2019}
\begin{equation}\label{scrivauto:52}
    \begin{aligned}
    \bar{\psi}_{j,\GKP}(s) \approx
    {\frac {\sqrt 2}{\pi^{1/4}}} \, 
    e^{-\frac{\kappa^2 s^2}{2}}\sum_{n=-\infty}^\infty
    e^{-\frac{(s-(2n+j)\sqrt{\pi})^2}{2\Delta^2}}
    \end{aligned}
    ,
\end{equation}
where the parameters that define the quality of a GKP state, the spike variance $\Delta^2$ and the envelope variance $\kappa^{-2}$~\cite{GKP}, are related to the damping factor by 
    \begin{equation}
        \Delta^2 = \kappa^2  = \beta.
    \end{equation}
Figure~\ref{scrivauto:53} shows $\ket{\bar{0}_\GKP}$ (solid) and $\ket{\bar{1}_\GKP}$ (dashed) for $\beta = 13.8 \times 10^{-3}$ corresponding to GKP squeezing of $s_\GKP=18.6$ dB by the relation: $s_\GKP= -10 \log_{10}(\Delta^{2}).$\footnote{Note that $s_\GKP < 0$ is squeezed and $s_\GKP > 0$ is anti-squeezed (with respect to vacuum).}

\subsection{Teleportation-based GKP error correction} \label{sec:GKPteleport}

The goal of GKP error correction is to eliminate CV-level noise that has corrupted an encoded state through a procedure that projects the noisy state back into the GKP codespace. Sources of noise include displacements \cite{GKP}, thermal noise including pure loss \cite{Noh2019}, dephasing \cite{Nielsen2010}, and inherent finite-squeezing noise that is particularly relevant in CVCS settings \cite{Nick2014}. In Steane-style GKP error correction, the noisy encoded state is coupled to a GKP ancilla, which is measured in the $q$ basis (using homodyne detection). The process is repeated with another GKP ancilla for the $p$ basis. The two measurement outcomes (referred to together as the syndrome) are fed into a decoder that provides a recovery map consisting of shifts in position and momentum that, when actively performed, restore the GKP subspace and attempt to correct logical errors on the encoded qubit \cite{GKP,Glancy2006}. 

In CVCS settings, Gaussian finite-squeezing noise accumulates as an encoded state teleports from node to node~\cite{Alexander2014}. 
It was shown that fault tolerance in canonical CVCS quantum computing is possible by periodically performing Steane-style GKP error correction \cite{Nick2014}.
However, canonical CVCSs and associated GKP error correction require two-mode entangling operations of the form $e^{ig\op{q} \otimes \op{q}}$, which are hard to realize because they require active squeezing. Experimentally accessible CVCSs are those used in macronode-CVCS QC, where the coupling requires only passive components---beam splitters and phase delays. 

We show here that an alternate method, known as Knill-style CV error correction, proceeds by teleporting the input state through two encoded ancillae---typically an encoded Bell pair~\cite{Baragiola2019}---performing the recovery map automatically. The syndrome is used purely to determine the likelihood and type of logical-level error \cite{Grimsmo2020}. In fact, pre-measuring the $\ket{0_\GKP}$ ancillae used for Steane-style error correction in the canonical CVCS construction of Ref.~\cite{Nick2014} (and performing shifts) accordingly embeds them directly inline as a 2-qubit GKP cluster state. Teleporting through this inline resource state may then be interpreted as Knill-style error correction. This enables two different interpretations of the same physical setup from Ref.~\cite{Nick2014}, 
    \begin{equation}
    \begin{split}\label{canonicalCVCS}
    \centering
    \includegraphics[width=.8\columnwidth]{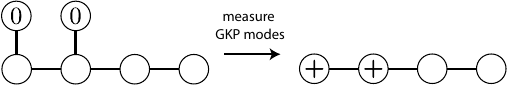}
    \end{split}\, ,
    \end{equation}
either Steane-style (left) or Knill-style (right) error correction.
Here, empty circles are modes in squeezed momentum states, non-empty circles are modes in GKP states, and lines between modes are $e^{i\op{q} \otimes \op{q}}$ gates.

Importantly, we show here that Knill-style teleportation-based error correction is directly compatible with macronode-based protocols, which use experimentally accessible CVCSs. Further, we show that the GKP states for error correction need not be placed adjacent to one another in the cluster state, a flexibility that is crucial for experimental settings with probablistically generated GKP states.

\subsubsection{GKP error correction on the macronode wire}
\label{macronodeGKPEC}
To demonstrate how the Kraus operator in Eq.~\eqref{genkrausdamp} can be used for GKP error correction on the macronode wire, we first consider the ideal case, $\beta=0$, with qunaught states for both $\ket{\psi}$ and $\ket{\phi}$. 
Using the qunaught wavefunctions in the expression for the teleported gate, Eq.~\eqref{EPRbounce}, gives
	\begin{equation}  \label{GKPAoperator_step1}
		\bounceEPR{\varnothing, \varnothing}{} = \iint d^2\alpha \, \Sha_{\sqrt{2 \pi}}(\alpha_R) \Sha_{\sqrt{2\pi}}(\alpha_I) \op D(\alpha) \, .
	\end{equation}
This expression can be brought into a convenient form by decomposing the displacement operator into position and momentum shifts with Eq.~\eqref{Ddecompose}, recognizing that the resulting phase is trivial for qunaught wavefunctions, and then rewriting the two resulting integrals using the Fourier relations,
    \begin{subequations}
	\begin{align}
		\psi(\op{q}) = &  \frac{1}{\sqrt{2\pi}} \int ds \, \mathcal{F}[\psi](s) \op{Z}(s) \\
		\mathcal{F}[\psi](\op{p}) &= \frac{1}{\sqrt{2\pi}} \int ds \, \psi(s) \op{X}(s) \, .
	\end{align}
    \end{subequations}
After these manipulations, the operator in  Eq.~\eqref{GKPAoperator_step1} is revealed to be the square-lattice GKP projector,
	\begin{subequations} \label{GKPAoperator}
	\begin{align}
		\bounceEPR{\varnothing, \varnothing}{} 
		    &= \pi\sqrt{2}\, \Sha_{\sqrt{\pi}}(\op{p}) \Sha_{\sqrt{\pi}}(\op{q}) \\
		    &= \pi\sqrt{2}\, \Sha_{\sqrt{\pi}}(\op{q}) \Sha_{\sqrt{\pi}}(\op{p})  \\
		    &= \sqrt{\frac{\pi}{2}}\, \GKPproj \, .
	\end{align}
    \end{subequations}
The final line is shown by expanding either of the previous lines using $\Sha_{T}(\op{q}) = \sqrt{T}\sum_n \delta(\op{q}-nT)$, where 
    \begin{equation}
        \delta(\op{q}-nT) = \outprodsubsub{nT}{nT}{q}{q}\, ,
    \end{equation}
and an analogous expansion for $\Sha_{T}(\op{p})$. Then, with $T = \sqrt\pi$, we see that
\begin{align}
    \Sha_{\sqrt\pi}(\op q)\Sha_{\sqrt\pi}(\op p)
&=
    \frac{1}{2\sqrt{\pi}} \GKPproj
    \,,
\end{align}
that is, these operators combine to form the GKP projector~\cite{Baragiola2019} (up to an overall factor). The circuit diagram for Eq.~\eqref{GKPAoperator} is
\begin{equation} \label{GKPBellpaircircuit}
\begin{split}
    \Qcircuit @C=1em @R=1.1em 
    {
 &\bsbal{1}  &\rstick{\ket{\qunaught}} \qw &&&&& \qw & \ar @{-} [dr(0.5)] \qw & \\
 & \qw &\rstick{\ket{\qunaught}} \qw &&&\ustick{\raisebox{.2em}{$=$}}&&       \gate{ \frac{1}{\sqrt{2}} \Pi_\GKP} & \ar @{-} [ur(0.5)] \qw &
		}
     \, \raisebox{-1em}{.}
\end{split}
\end{equation}

Inserting Eq.~\eqref{GKPAoperator} into the general form for the Kraus operator, Eq.~\eqref{genkraus}, with $\beta=0$ gives the Kraus operator for macronode-based GKP error correction,\footnote{The factor preceding the GKP projector is a consequence of our chosen normalization for ideal GKP states, Eqs.~\eqref{GKPqbasis} and~\eqref{GKPpbasis}.}
    \begin{equation} \label{genkrausGKP}
        \op{K}_\GKP(m_a,m_b) = \frac{1}{\sqrt{2\pi}}  \GKPproj \op{D}(\mu) \op{V}(\theta_a,\theta_b) . 
    \end{equation}
This expression shows that after the measurement-basis-dependent Gaussian unitary $\op{V}(\theta_a ,\theta_b)$ [Eq.~\eqref{Vgate}] and outcome-dependent displacement $\op{D}(\mu)$, the state is projected into the GKP subspace. The whole gadget can be interpreted as teleporting through a square-lattice GKP Bell pair \cite{Baragiola2019}, Eq.~\eqref{gkpbellpair}, with additional damping when the ancillae are imperfect, as we show below.

\subsubsection{Approximate qunaught ancillae in the Kraus state}
In physical settings, the qunaught ancillae contain finite-squeezing noise, which we model as damped ideal qunaught states,
    \begin{equation} \label{approxqunaught}
          \ket{\bar{\qunaught}} \coloneqq \frac{1}{ \sqrt{ \mathcal{N}_\qunaught} } \dampop{\beta} \ket{\qunaught}  \, ,
    \end{equation}
with normalization factor,
    \begin{equation}
        \mathcal{N}_{\qunaught} \coloneqq \bra{\qunaught} \dampop{2\beta} \ket{\qunaught}  \, ,
    \end{equation}    
using the damping operator $\dampop{\beta}$ in Eq.~\eqref{damping}.    
We then use the circuit identities above to write the beam splitter entangled state as joint damping on a GKP Bell pair,
	\begin{equation} \label{caseAB}
	\begin{split}
     \Qcircuit @C=0.45cm @R=0.3cm 
    {
         &\bsbal{1} &\rstick{\ket{\bar{\qunaught}}} \qw &&                               &&\gate{ \mathcal{N}^{-1/2}_\qunaught\dampgate} &\bsbal{1} &\rstick{\ket{\qunaught}} \qw &\\
         &\qw       &\rstick{\ket{\bar{\qunaught}}} \qw && \ustick{\raisebox{.5em}{$=$}} &&\gate{ \mathcal{N}^{-1/2}_\qunaught \dampgate} &\qw        &\rstick{\ket{\qunaught}} \qw &      
    }
    \end{split}
    \end{equation}
The macronode Kraus operator for this situation is obtained by making the replacement ,
    \begin{equation} \label{approxGKPcase}
        \op{\Pi}_\GKP \rightarrow \frac{1}{ \mathcal{N}_{\qunaught} } \dampop{\beta} \GKPproj \dampop{\beta} = \frac{1}{ \mathcal{N}_{\qunaught} } \approxGKPproj \, ,
    \end{equation}
in the ideal Kraus operator $\op{K}_\GKP(m_a,m_b)$, Eq.~\eqref{genkrausGKP}, and $\approxGKPproj$ is given in Eq.~\eqref{dampGKPprojector}.

The GKP error-correction Kraus operator transforms an arbitrary input pure state $\ket{\Psi}$ traversing the teleportation gadget via the Kraus map in Eq.~\eqref{Krausmapdef}. For damped qunaught ancillae---that is, using the Kraus operator when Eq.~\eqref{approxGKPcase} is used with Eq.~\eqref{genkrausGKP}---results in the GKP-encoded state,
    \begin{equation}
        \ket{\Psi_\GKP} = c_0 \ket{\bar{0}_\GKP} + c_1 \ket{\bar{1}_\GKP}
    \end{equation}
whose qubit coefficients are given by
    \begin{align} \label{GKPcoefs}
        c_j = \frac{ 1 }{ \sqrt{ \Pr(m_a,m_b)} }  \frac{ \mathcal{N}_{j,\GKP} }{\sqrt{2\pi} \mathcal{N}_\qunaught} \bra{\bar{j}_\GKP} \op{V}(\theta_a,\theta_b) \op{D}(\mu) \ket{\Psi} \, ,
    \end{align}
and the joint probability of obtaining the outcomes, $\Pr(m_a,m_b)$, is given by Eq.~\eqref{jointprob}. The input state is actively projected into the GKP subspace, and the syndrome information (the two homodyne measurement outcomes) is used by a GKP decoder to determine the likelihood of a logical error. Effectively, this is a projection of excess CV noise into GKP-logical noise. The benefit of a teleportation-based approach to error correction is that the output state is already in the GKP subspace, meaning that corrections are purely logical operations (i.e.,~shifts by integer multiples of~$\sqrt\pi$). This differs from Steane-style error correction, in which the corrections involve small displacements to realign the resulting state with the GKP grid~\cite{GKP,Glancy2006}. For the case of no damping, $\beta \rightarrow 0$, the ratio of normalization factors in Eq.~\eqref{GKPcoefs} disappears, and the projection is onto the ideal, orthogonal GKP states $\ket{j_\GKP}$.

\subsubsection{Partial GKP error correction} 
From Eqs.~\eqref{GKPAoperator}, we see that using qunaught states for both ancillae yields a separable operation: a projection in $\sqrt{\pi}$-periodic position followed by a projection in $\sqrt{\pi}$-periodic momentum that together form the GKP codespace projector. Interpreted logically, each of these projections corresponds to measurement and recovery of one of the two GKP stabilizers. 
Here, we show that each of these projections can take place separately, such that half of GKP error correction can occur at one macronode and the other half at another macronode---\emph{i.e.} the periodic $q$ projection and then, later, the periodic $p$ projection. This is particularly useful for practical implementations of GKP error correction using probabilistic sources of GKP states.

\begin{figure}[t!]
\centering
\includegraphics[width=1\columnwidth]{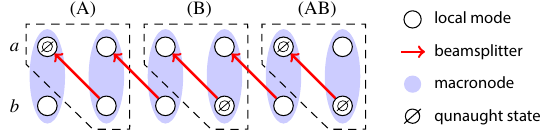}

  \begin{center}
    \label{scrivauto:108}
    \bgroup
\def\arraystretch{1.7}
   \begin{tabular}{m{.15\columnwidth}|m{.25\columnwidth}|m{.25\columnwidth}|m{.25\columnwidth}|}
\cline{2-4}
 & \textbf{Case (A)} & \textbf{Case (B)} & \textbf{Case (AB)} \\ \hline
\multicolumn{1}{|l|}{$\phi(\alpha_R)$} & $\Sha_{\sqrt{2\pi}}(\alpha_R)$ & $\delta(\alpha_R)$  &    $\Sha_{\sqrt{2\pi}}(\alpha_R)$      \\ \hline
\multicolumn{1}{|l|}{$\widetilde \psi(\alpha_I)$}        & $\delta(\alpha_I)$            & $\Sha_{\sqrt{2\pi}}(\alpha_I)$ &  $\Sha_{\sqrt{2\pi}}(\alpha_I)$ \\ \hline
\end{tabular}
\egroup
\end{center}
\caption{Potential qunaught sites for GKP error correction. A macronode wire is depicted with the possible locations for qunaught states at either mode $a$ (A) or $b$ (B) within a teleportation gadget. When qunaught states are placed at both modes, they comprise a GKP Bell pair shared between adjacent macronodes (AB). The table shows the momentum and position wave functions to be used in the Kraus operator, Eq.~\eqref{genkraus}, for each case.}\label{GKPsites}
\end{figure}

We consider two new cases, indicated graphically in Fig.~\ref{GKPsites}: (A) only mode $a$ is prepared in a qunaught state, and (B) only mode $b$ is prepared in a qunaught state. The non-qunaught mode is either a momentum-squeezed~(A) or a position-squeezed~(B) state following the standard macronode procedure. The case (AB) where both ancillae modes are qunaught states is considered above in Sec.~\ref{macronodeGKPEC}. 
The respective wave functions for the ancillae, $\tilde \psi(\alpha_I)$ and $\phi(\alpha_R)$, to be used in Eq.~\eqref{EPRbounce} for each case are given in the table in Fig.~\ref{GKPsites}.

Case (A) gives
\begin{equation}
    \bounceEPR{\psi, \phi}{\GKP,A}  = 2^{\frac 1 4} \sqrt\pi \Sha_{\sqrt\pi}(\op p)\,.
\end{equation}
Using Eq.~\eqref{ancillacircuit}, we get the following circuit identity:
\begin{equation} \label{GKPcircuitA}
\begin{split}
    \Qcircuit @C=1em @R=1.1em 
    {
 &\bsbal{1}  &\rstick{\pket 0} \qw &&&&& \qw & \ar @{-} [dr(0.5)] \qw & \\
 & \qw &\rstick{\ket{\qunaught}} \qw &&&\ustick{\raisebox{.2em}{$=$}}&&       \gate{ 2^{\frac 1 4} \Sha_{\sqrt\pi}(\op p)} & \ar @{-} [ur(0.5)] \qw &
		}
     \, \raisebox{-1em}{.}
\end{split}
\end{equation}
Using Eq.~\eqref{genkraus}, this yields a Kraus operator that performs partial GKP error correction in the $p$ quadrature,
    \begin{equation} \label{caseAkraus}
        \op{K}_{\GKP,A}(m_a,m_b) 
            = \frac{2^{\frac{1}{4}}}{\sqrt{\pi}} \Sha_{\sqrt{\pi}}(\op{p}) \op{D}(\mu) \op{V}(\theta_a,\theta_b) 
            \,.
    \end{equation}

Case (B) is almost the same (with $\op q \leftrightarrow \op p$). Plugging the wave functions for case (B) into Eq.~\eqref{EPRbounce} gives
\begin{equation}
    \bounceEPR{\psi, \phi}{\GKP,B}  = 2^{\frac 1 4} \sqrt\pi \Sha_{\sqrt\pi}(\op q)\,,
\end{equation}
whose circuit is
\begin{equation} \label{GKPcircuitB}
\begin{split}
    \Qcircuit @C=1em @R=1.1em 
    {
 &\bsbal{1}  &\rstick{\ket \qunaught} \qw &&&&& \qw & \ar @{-} [dr(0.5)] \qw & \\
 & \qw &\rstick{\qket 0} \qw &&&\ustick{\raisebox{.2em}{$=$}}&&       \gate{ 2^{\frac 1 4} \Sha_{\sqrt\pi}(\op q)} & \ar @{-} [ur(0.5)] \qw &
		}
     \, \raisebox{-1em}{.}
\end{split}
\end{equation}
This yields a Kraus operator that performs partial GKP error correction in the $q$ quadrature,
    \begin{equation} \label{caseBkraus}
        \op{K}_{\GKP,B}(m_a,m_b) 
        = \frac{2^{\frac{1}{4}}}{\sqrt{\pi}}  \Sha_{\sqrt{\pi}}(\op{q}) \op{D}(\mu) \op{V}(\theta_a,\theta_b) \, .
    \end{equation}

Each Kraus operator acts as a ``filter'' on the input state in either $\sqrt{\pi}$-periodic momentum or position, respectively. For measurement bases chosen such that $\op{V}(\theta_a,\theta_b) = \op{I}$ and displacements $\op{D}(\mu)$ at each macronode accounted for by active corrective shifts, the two partial GKP error corrections can occur at separated macronodes. For ideal quadrature eigenstates and qunaught states, \emph{i.e.}, neither contains finite-squeezing noise, the result is ideal GKP error correction.

We model physical states at the ancillae with damped qunaught states, Eq.~\eqref{approxqunaught}, and squeezed states described by Eq.~\eqref{sqzstate} as approximate quadrature eigenstates. For both types of state, we assume that the damping is equal so the damping operators can be extracted and the remaining qunaught and quadrature eigenstates treated as ideal.

The circuit identity for case (A) is
	\begin{equation} \label{caseA}
	\begin{split}
 \Qcircuit @C=0.5cm @R=0.3cm 
    {
         & \bsbal{1} &\rstick{\pket{0; \zeta} }       \qw &&&                               &&\gate{ \mathcal{N}_\zeta^{-1/2} \dampgate}     &\bsbal{1} &\rstick{\pket{0}}       \qw &\\
         &\qw        &\rstick{\ket{\bar{\qunaught}} } \qw &&& \ustick{\raisebox{.5em}{$=$}} &&\gate{ \mathcal{N}_\qunaught^{-1/2} \dampgate} &\qw      &\rstick{\ket{\qunaught}}\qw & 
    }
    \end{split}
    \end{equation}
and the Kraus operator performs partial GKP error correction in the $p$ quadrature,
    \begin{equation}  
         \Sha_{\sqrt{\pi}}(\op{p}) \rightarrow 
             \frac{1}{ \sqrt{ \mathcal{N}_{\qunaught}\mathcal{N}_{\zeta}} } \dampop{\beta} \Sha_{\sqrt{\pi}}(\op{p}) \dampop{\beta}
            \,.
    \end{equation}
Similarly, for case (B), the circuit identity is
	\begin{equation} \label{caseB}
	\begin{split}
    \Qcircuit @C=0.5cm @R=0.3cm 
        {
         &\bsbal{1} &\rstick{\ket{\bar{\qunaught}}} \qw &&&                               &&\gate{ \mathcal{N}_\qunaught^{-1/2}\dampgate} &\bsbal{1} &\rstick{\ket{\qunaught}} \qw &\\
         &\qw        &\rstick{\qket{0; \zeta} }     \qw &&& \ustick{\raisebox{.5em}{$=$}} &&\gate{ \mathcal{N}_\zeta^{-1/2}\dampgate}     &\qw       &\rstick{\qket{0}}\qw & 
         }
    \end{split}
    \end{equation}
and the associated Kraus operator performs partial GKP error correction in the $q$ quadrature,
    \begin{equation} 
         \Sha_{\sqrt{\pi}}(\op{q}) \rightarrow 
             \frac{1}{\sqrt{ \mathcal{N}_{\qunaught}\mathcal{N}_{\zeta}} } \dampop{\beta} \Sha_{\sqrt{\pi}}(\op{q}) \dampop{\beta}     \, .
    \end{equation}
The benefit of performing full GKP error correction utilising a GKP Bell pair is simply that there is no buildup of finite-squeezing noise (and potential external noise) between the two partial GKP error corrections.

\section{Application: GKP error correction in macronode lattices}
\label{multiD}

Above we analyzed the teleportation-gadget Kraus operator when one or more modes are prepared in the GKP qunaught state $\ket{\qunaught}$.  By combining this operator with the usual measurement-based evolution with Gaussian resource states, we can describe a hybrid resource state made up of both Gaussian and non-Gaussian parts. These results can be generalized to other CV cluster states made from two-mode CV cluster states and 50:50 beam splitters.

All multidimensional macronode CV cluster states proposed in Refs.~\cite{Menicucci2011a, Wang2014, Alexander2016, Asavanant2019, Larsen2019, wu2020quantum} can be constructed by arranging two-mode cluster states in a higher dimensional geometry and then producing a fully connected resource by applying a local constant-depth circuit of 50:50 beam splitters. In all of these cases, the resulting lattices contain embedded macronode wires that are used to implement single-mode gates. Said another way, each such cluster state is equivalent to a collection of macronode wires coupled via additional 50:50 beam splitters. By choosing homodyne measurements appropriately, some of these beam splitters can be ``cancelled out,'' producing local regions in the lattice that are literally equivalent to a macronode wire. Details can be found in the references above. The upshot is that teleportation-based GKP error correction, described in Sec.~\ref{sec:GKPteleport}, has applications in all of the higher dimensional macronode-based architectures known to date. This combines the single-mode Gaussian gates and GKP error correction of a single macronode wire with the ability to implement multimode Gaussian unitary gates, resulting in a universal set of resources~\cite{Baragiola2019} via passive linear optics and offline preparation of squeezed vacuum and GKP $\ket{\qunaught}$ states.

\section{Conclusion}

We have presented the Kraus operator for CV gate teleportation based on teleporting through a class of entangled states that are prepared on a beam splitter. The teleported gate depends on the ancilla states at the beam splitter, which themselves can be non-Gaussian, enabling implementation of teleported non-Gaussian operations.
 Our analysis focuses on the macronode wire, but it is directly mappable onto higher-dimensional macronode CV cluster states because all known macronode cluster states are equivalent to a collection of macronode wires coupled via additional 50:50 beam splitters.
 
 We use the derived formalism to propose a teleportation-based scheme for GKP error-correction requiring only constant-depth linear optics to implement, removing the requirement of previous work on active squeezing operations. Further, our GKP error-correction scheme is compatible with nondeterministic state preparation \cite{Bourassa2020}. This simplifies the implementation of practical GKP error correction---the final ingredient needed by CV cluster state architectures for universal and fault-tolerant computation---bringing it in line with the experimentally available resources.

\acknowledgments
This work was supported by the Australian Research Council Centre of Excellence for Quantum Computation and Communication Technology (Project No. CE170100012).
R.N.A.\ is supported by National Science Foundation
Award No. PHY-1630114.

\appendix
\numberwithin{equation}{section}
\renewcommand{\theequation}{\thesection\arabic{equation}}

\section{Entangling quadrature eigenstates on a beam splitter}\label{appendix:beamsplitterstates}

We start with the LDU and UDL decompositions of a rotation matrix $\mat{R}(\theta)$ valid for $\theta \neq n \pi/2$ for odd $n$, 
\begin{subequations} \label{genrotationDecomp}
\begin{align} 
    \mat{R}(\theta) &=
    \begin{pmatrix}
         \cos{\theta} & - \sin{\theta}
    \\
         \sin{\theta} &  \cos{\theta}
    \end{pmatrix}
    \nonumber\\
&= \label{genrotationDecomp1}
    \begin{pmatrix}
        1 & 0
    \\
        \tan{\theta} & 1
    \end{pmatrix}
    \begin{pmatrix}
        \cos{\theta} & 0
    \\
        0 & \sec{\theta}
    \end{pmatrix}
    \begin{pmatrix}
        1 & -\tan{\theta}
    \\
        0 & 1
    \end{pmatrix}
 \\
&= \label{genrotationDecomp2}
    \begin{pmatrix}
        1 & -\tan{\theta}
    \\
        0 & 1
    \end{pmatrix}
    \begin{pmatrix}
        \sec{\theta} & 0
    \\
        0 & \cos{\theta}
    \end{pmatrix}
    \begin{pmatrix}
        1 & 0
    \\
        \tan{\theta} & 1
    \end{pmatrix}
    .
\end{align}
    \end{subequations}
Each decomposes $\mat{R}(\theta)$ into a product of symplectic matrices describing a shear, squeezing, and another shear.

Decomposing the symplectic representation of the Heisenberg action of the beam splitter $\bsop_{jk}$ that acts on the vector of quadratures~$(\op q_1, \op q_2, \op p_1, \op p_2)^\tp$~\cite{alexander2016flexible}, namely
\begin{align}
\label{eq:BSsymplecticdecomp}
&    
    \begin{pmatrix}
    \mat R(\tfrac{\pi}{4}) & \mat 0 \\ \mat 0 & \mat R(\tfrac{\pi}{4})
    \end{pmatrix}
\nonumber \\
&=
    \begin{pmatrix}
        1 & 0 & 0 & 0 \\
        1 & 1 & 0 & 0 \\
        0 & 0 & 1 & -1 \\
        0 & 0 & 0 & 1
    \end{pmatrix}
    \begin{pmatrix}
        \frac{1}{\sqrt{2}} & 0 & 0 & 0 \\
        0 & \sqrt 2 & 0 & 0 \\
        0 & 0 & \sqrt 2 & 0 \\
        0 & 0 & 0 & \frac{1}{\sqrt{2}} 
    \end{pmatrix}
    \begin{pmatrix}
        1 & -1 & 0 & 0 \\
        0 & 1 & 0 & 0 \\
        0 & 0 & 1 & 0 \\
        0 & 0 & 1 & 1
    \end{pmatrix}
    ,
\end{align}
yields, up to an overall phase, a unitary decomposition of the beam splitter:
\begin{equation}
\label{eq:BSunitarydecomp}
    \bsop_{12} = e^{-i \op q_1 \otimes \op p_2}
    [\op S_1^\dag({\sqrt{2}}) \otimes \op S_2({\sqrt{2}})]
    e^{i \op p_1  \otimes\op q_2}
    .
\end{equation}
We can write this as a circuit identity (again, up to an overall phase),
    \begin{equation}
\label{eq:BScircuitdecomp}
    \begin{split}
         \Qcircuit @C=1em @R=2.5em @! 
         {
         	& \bsbal{1} & \qw \\
         	& \qw       & \qw
  		  }
\quad \raisebox{-1.3em}{$=$} \quad
    \Qcircuit @C=1em @R=1em {
         	& \ctrl{1} & \gate{S^\dag({\sqrt{2}})} & \targ & \qw\\
         	& \targ & \gate{S(\sqrt{2})} & \ctrlo{-1} & \qw
  		  }
    \end{split} \, ,
	\end{equation}
where the two-mode gates are each a CV controlled-$X$ gate, Eq.~\eqref{controlledX}, of weight~$g=\pm 1$, with the open circle indicating weight~$-1$.\footnote{Again, notice that the right-to-left circuit convention makes such conversions straightforward: Eqs.~\eqref{eq:BSsymplecticdecomp}, \eqref{eq:BSunitarydecomp}, and \eqref{eq:BScircuitdecomp} are all ordered the same way.}

Consider momentum and position eigenstates, $\ket{t}_{p}$ and $\ket{s}_{q}$, coupled by a beam splitter. Using the above decomposition, the resulting entangled state is
\begin{align}
    \bsop_{12} \ket{t}_{p_1} \otimes \ket{s}_{q_2}
&=
    e^{ist} e^{-i \op q_1 \otimes \op p_2} \op S_1^\dag({\sqrt{2}}) \ket{t}_{p_1} \otimes \op S_2({\sqrt{2}}) \ket{s}_{q_2}
\nonumber \\
&=
    \sqrt{2} e^{ist} e^{-i \op q_1  \otimes\op p_2} \ket{\sqrt{2}t}_{p_1} \otimes \ket{\sqrt{2}s}_{q_2}
\nonumber \\
&=
     \sqrt{2} e^{ist} \ket{\EPR(\sqrt{2}s,\sqrt{2}t)}\, , \label{BSalmostthere_appendix} 
\end{align}
again up to a fixed phase. The norm scaling by~$\sqrt{2}$ arises from the action of the squeezing operator on a quadrature eigenstate:
\begin{subequations} \label{squeezeeffect}
\begin{align}
    \op S(\zeta) \qket s
&=
    \abs\zeta^{1/2} \qket {\zeta s}\,,
\\
    \op S(\zeta) \pket t
&=
    \abs\zeta^{-1/2} \pket {\zeta^{-1}t}\,,
\end{align}
\end{subequations}
where $\op S(\zeta)$ is defined in Eq.~\eqref{squeezingop}.
The prefactor ensures that inner products are preserved and that projectors onto squeezed quadrature eigenstates resolve the single-mode identity when integrated over~$s$ (or $t$).

\section{Derivation of the gates for rotated homodyne measurements} \label{appendix:Vop}

The results of the previous Appendix can be generalized to the case where eigenstates of rotated quadratures, such as those in Eq.~\eqref{scrivauto:43}, are coupled on a beam splitter. The resulting state is equivalent to a set of single-mode Gaussian operations and a displacement acting on an EPR state. Important for our purposes is the fact that the Hermitian conjugate of this state gives a description of the two-mode entangled homodyne measurement in the macronode teleportation gadget, Eq.~\eqref{scrivauto:77}. The operations on the EPR state are indeed those that allow universal Gaussian quantum computation with standard (Gaussian) macronode cluster states: the operator $\op{V}(\theta_a, \theta_b)$ in Eq.~\eqref{Vgate}, as well as a displacement $\op{D}(\mu)$. Here, we give a derivation of these operations with specific attention to how they arise in this measurement. In this context, note that they arise solely from the choice of homodyne measurement angles and outcomes and are distinct from the gate teleported by the Kraus state (although we use the same formalism in its derivation). 

Rotated momentum and position quadratures, 
    \begin{align}
        \op{p}_{\theta} &\coloneqq \op{R}^\dagger(\theta) \op{p} \op{R}(\theta) =  \sin \theta \op q  + \cos \theta\op p  \, , \\
        \op{q}_{\theta} &\coloneqq \op{R}^\dagger(\theta) \op{q} \op{R}(\theta) =  \cos \theta \op q - \sin \theta \op p  \, ,
    \end{align}
have respective eigenstates $\op{p}_{\theta}\ketsub{t}{p_{\theta}} = t \ketsub{t}{p_{\theta}}$ and $\op{q}_{\theta}\ketsub{s}{q_{\theta}} = s \ketsub{s}{q_{\theta}}$. We express these states in a convenient way using the decompositions of the rotation operator in Eqs.~\eqref{genrotationDecomp},
\begin{align}
    \ketsub{t}{p_{\theta}} 
        &\coloneqq \op R^\dagger(\theta) \op{Z}(t) \pket{0}\\
        & = \op D^{\dag}\big(-\tfrac{is}{\sqrt{2}}e^{-i\theta}\big) \op R^\dagger(\theta) \pket{0}\\
        &= \sqrt{|\sec{\theta}|} \op D^{\dag}\big(-\tfrac{it}{\sqrt{2}}e^{-i\theta} \big) \op P(-\tan{\theta})\pket{0}\, ,
        \label{pthetaket} \\
    \ketsub{s}{q_{\theta}} 
        &\coloneqq \op R^\dagger(\theta) \op{X}(s) \qket{0}\\
        &= \op D^{\dag}\big (-\tfrac{s}{\sqrt{2}}e^{-i\theta}\big) \op R^\dagger(\theta) \qket{0} \\
        &= \sqrt{|\sec{\theta}|} \op D^{\dag} \big(-\tfrac{s}{\sqrt{2}}e^{-i\theta}\big) \op P_p(\tan{\theta})\qket{0}. \label{qthetaket}
\end{align}
In the final line of each expression are unitary shear operators, defined respectively as
    \begin{align}
        \op P(\sigma)  &\coloneqq e^{\frac{i \sigma}{2} \op{q}^2} \\
        \op P_p(\sigma) &\coloneqq e^{-\frac{i \sigma}{2} \op{p}^2}.
    \end{align}
Note that each shear does nothing to a particular 0-eigenstate, $\op P(\sigma)\qket{0} = \qket{0}$ and $\op P_p(\sigma)\pket{0} = \pket{0}$, which allowed for the simplifcations in Eqs.~\eqref{pthetaket} and \eqref{qthetaket}. The factor in front of each expression arises from the action of squeezing operators on quadrature eigenstates, Eq.~\eqref{squeezeeffect}.

We now focus on the rotated measurements in the macronode teleportation gadget, Eq.~\eqref{scrivauto:77}, that are written as a projection onto a two-mode entangled state,
\begin{equation} \label{twomodemeasurement}
\begin{split}
\centering
    \Qcircuit @C=1.5em @R=2em 
    {
	&&\lstick{\brasub{m_a}{p_{\theta_a}}}   & \bsbal{1} & \qw &   \\
	&&\lstick{\brasub{m_b}{p_{\theta_b}}}  & \qw        & \qw & \\
	} 
	\end{split} \, .
\end{equation}
As the ultimate goal is to write this expression in terms of an EPR state of the form in Eq.~\eqref{beamsplitterquadstates} , we include in the rotation angle on the top mode an additional $\pi/2$, so that it describes a rotation from the $q$ axis of that mode,
    \begin{equation} \label{thetaaprimed}
        \theta_a' \coloneqq \theta_a - \pi/2 \, .
    \end{equation}
For convenience, we take the Hermitian conjugate of the expression (which reverses the direction of the beam splitter), to get
\begin{equation} \label{conjugatemeasurementstate}
        \begin{split}
		 \Qcircuit @C=1.5em @R=1em {
		 	&\qw &\gate{R^{\dag}(\theta_a')}      &\rstick{\ketsub{m_a}{q}} \qw &\\
         	&\bsbal{-1} & \gate{R^{\dag}(\theta_b)} &\rstick{\ketsub{m_b}{p}} \qw &
  		  } \, 
  		\end{split} \quad .
	\end{equation}
We split each phase delay operator into symmetric and anti-symmetric parts, $\op{R}(\theta) = \op{R}(\theta_+) \op{R}(\theta_-)$, with
    \begin{equation} \label{thetapm}
        \theta_{\pm}' \coloneqq \frac{\theta_a' \pm \theta_b}{2} \, ,
    \end{equation}
since the common delays (rotations) commute with any beam splitter, 
	\begin{equation}\label{rotatedinputs}
	\begin{split}
	\Qcircuit @C=0.3cm @R=0.3cm {
		 	&\qw        &\gate{R^{\dag}(\theta_a')} & \qw  \\
         	&\bsbal{-1} & \gate{R^{\dag}(\theta_b)} & \qw 
  		  }
  		  \raisebox{-1.3em}{=}
    \Qcircuit @C=0.3cm @R=0.3cm {
        &\gate{R^{\dag}(\theta_+')} &\qw &\gate{R^{\dag}(\theta_-')}       &\qw  \\
        &\gate{ R^{\dag}(\theta_+')} &\bsbal{-1}&\gate{R^{\dag}(-\theta_-')} & \qw 
         }
	\end{split}
    \end{equation}

Finally, we extract the displacements from each state in Eq.~\eqref{conjugatemeasurementstate}, commute them past the beam splitter, and use Eqs.~\eqref{pthetaket} and \eqref{qthetaket} to express the remaining rotated quadrature eigenstates as sheared eigenstates,
\begin{equation}\label{shearedinputs}
    \Qcircuit @C=0.3cm @R=0.3cm {
     &\gate{R^{\dag}(\theta_+')} &\gate{D^{\dag}(-\beta_+)}&\qw &\gate{\sqrt{|\sec{\theta_-'}|}P_p(\tan{\theta_-'})}       &\rstick{\qket{0}}\qw &\\
        &\gate{ R^{\dag}(\theta_+')} & \gate{D^{\dag}(\beta_-)}&\bsbal{-1}&\gate{\sqrt{|\sec{\theta_-'}|}P(\tan{\theta_-'})} &\rstick{\pket{0}} \qw &
         }
    \end{equation}
where the displacement amplitudes are
    \begin{equation}
        \beta_{\pm} \coloneqq \frac{1}{2}(m_ae^{-i\theta_-'} \pm im_be^{i\theta_-'})\, .
    \end{equation}

We now recognize that the right-hand side of the circuit is a Kraus state of the form in Eq.~\eqref{ancillacircuit} with the ancillae given by
\begin{align}
    \ket{\psi}&=\sqrt{|\sec{\theta_-'}|}
    \op P(\tan{\theta_-'})\pket{0}\, ,\\
    \ket{\phi}&=\sqrt{|\sec{\theta_-'}|} \op P_p(\tan{\theta_-'})\qket{0}\, .
\end{align}
We can now directly evaluate the teleported gate that they enact using Eq.~\eqref{EPRbounce}. This requires obtaining wavefunctions for $\ket{\psi}$ and $\ket{\phi}$. The following are easy to evaluate directly using the forms above:
\begin{align}
    \psi(s) \coloneqq \inprodsubsub{s}{\psi}{q}{}
&=
    \frac{1}{\sqrt{2\pi |\cos{\theta_-'}|}}
    e^{\frac{i}{2} (\tan{\theta_-'}) s^2}
    \, ,
\\
    \tilde \phi(t) \coloneqq \inprodsubsub{t}{\phi}{p}{}
&=
    \frac{1}{\sqrt{2\pi |\cos{\theta_-'}|}}
    e^{-\frac{i}{2} (\tan{\theta_-'}) t^2}
    \, .
\end{align}
Taking the appropriate Fourier transforms, Eq.~\eqref{Fourier}, we obtain the other two, which we need for Eq.~\eqref{EPRbounce}:
\begin{align}
    \tilde \psi(t) = \mathcal{F}[\psi] (t) &= \frac{1}{\sqrt{2\pi |\cos{\theta_-'}|}} e^{ \frac{i}{2} (\cot{\theta_-'}) t^2} \, ,\\
   \phi(s) =\mathcal{F}^{-1}\bigl[\tilde \phi \bigr] (s) &= \frac{1}{\sqrt{2\pi |\cos{\theta_-'}|}} e^{ -\frac{i}{2}(\cot{\theta_-'})s^2}\, .
\end{align}
Substituting these wavefunctions into the teleported gate, Eq.~\eqref{EPRbounce} ($t=\alpha_I, s=\alpha_R$), gives the Weyl representation of some operator $\op O$,
\begin{align}
    \op O =\frac{1}{2\pi |\cos{\theta_-'}|} \int d^2 \alpha \, e^{ -\frac{i}{2}(\cot{\theta_-'}) (\alpha_R^2-\alpha_I^2)}\op D(\alpha)
\end{align}
with characteristic function \cite{alessandro2005},
\begin{align}\label{glauberfunc}
    \Tr[\op O \op D^{\dag}(\alpha)] = \frac{1}{2|\cos{\theta_-'}|}e^{ -\frac{i}{2}(\cot{\theta_-'})  (\alpha_R^2-\alpha_I^2)} \, .
\end{align}
We now show that the rotated squeezing operator $\op O = \op R^{\dag}(\frac{\pi}{4}) \op S(\zeta) \op R(\frac{\pi}{4})$, yields the same characteristic function, and we find the squeezing $\zeta$ as a function of $\theta_-'$. 
Evaluation of this characteristic function,
\begin{equation}
   \Tr[\op R^{\dag}(\tfrac{\pi}{4}) \op S(\zeta) \op R(\tfrac{\pi}{4}) \op D^{\dag}(\alpha)]
   =\Tr[ \op S(\zeta) \op D^{\dag}(\alpha e^{i \frac{\pi}{4} })]\, ,
\end{equation}
with the trace taken in the position basis gives
\begin{equation}
    \int dt \, \qbra{t} \op S(\zeta) \op D^{\dag} \big(\alpha e^{i\frac{\pi}{4}} \big)\qket{t} 
    = \frac{\sqrt{|\zeta|}}{|1-\zeta|} e^{-\frac{i}{2}\frac{1+\zeta}{1-\zeta}(\alpha_R^2-\alpha_I^2)} \, .
\end{equation}
Since the functional form of this expression is the same as that in Eq.~\eqref{glauberfunc}, we can read off the relations between parameters,
\begin{align} \label{sfactor}
     \cot{\theta_-'}&=  \frac{1+\zeta}{1-\zeta} \quad \longrightarrow \quad \zeta = - \tan \big(\theta_-' - \tfrac{\pi}{4} \big)\, .
\end{align}
Thus, we have shown that, for the Kraus state on the right-hand side of the circuit in Eq.~\eqref{shearedinputs}, the teleported gate is a $\frac{\pi}{4}$-rotated squeezing operation. 

Combining this result with the remaining rotations and displacements, Eq.~\eqref{shearedinputs} becomes 
	\begin{equation} \label{}
	\resizebox{\columnwidth}{!}{
 \Qcircuit @C=1em @R=1.5em {     
                     &&\gate{R^{\dag}(\theta_+')} &\gate{D^{\dag}(-\beta_+)}&\gate{R^{\dag}(\frac{\pi}{4})} &\gate{ \frac{1}{\sqrt{\pi}} S(\zeta)}& \gate{R(\frac{\pi}{4})} & \ar @{-} [dr(0.5)] \qw  &\\
         &&\qw &\qw&\qw    &\gate{R^{\dag}(\theta_+')}    & \gate{D^{\dag}(\beta_-)}  & \ar @{-} [ur(0.5)] \qw  & 
    } \, \raisebox{-1.5em}{,}
    }
\end{equation}
Note that the teleported gate is on the top wire here due to the beam splitter orientation in the state, Eq.~\eqref{conjugatemeasurementstate}.
Bouncing all the operators to the top wire using rules described in Eqs.~\eqref{scrivauto:89} and \eqref{dispbounce}, and then combining the two displacements reduces this to
\begin{equation} \label{almostcompleteV}
 \Qcircuit @C=0.3cm @R=0.5cm {                                                    
          &\gate{R^{\dag}(\theta'_+ + \tfrac{\pi}{4})} &\gate{\frac{1}{\sqrt{\pi}} S(\zeta)} &\gate{R^{\dag}(\theta'_+ - \tfrac{\pi}{4})} &\gate{D^{\dag}(\mu)} &\ar @{-} [dr(0.5)] \qw  &\\
          &\qw                                         &\qw                              &\qw                                         &\qw                   &\ar @{-} [ur(0.5)] \qw  & 
    } \; \raisebox{-1em}{,}
\end{equation}
where the final displacement amplitude~$\mu$ can be derived by bouncing all operators to the top mode and commuting both displacements to the end, giving ${\mu = -e^{i\theta_+'}[\beta_+  \cosh{(-\ln{\zeta})}-i \beta_+^* \sinh{(-\ln{\zeta})}+\beta_-^*]}$. The phase~$e^{i\theta_+'}$ at the front arises from bouncing and then commuting the final~$\op R^\dag(\theta'_+)$ through all displacements, the $\beta_-^*$ results from bouncing the $\op D^\dag(\beta_-)$ to the top wire, and the remaining terms result from commuting $\op D^\dag(-\beta_+)$ through the rotated squeezing operator. In the end, $\mu$ evaluates to
    \begin{equation} \label{mu}
        \mu =- \frac{ m_a e^{i \theta_b} + m_b e^{i \theta_a } }{ \sin (\theta_a - \theta_b)}
        \,
        .
    \end{equation}

Writing Eq.~(\ref{almostcompleteV}) in terms of (combinations of) the original measurement angles using Eqs.~\eqref{thetaaprimed} and \eqref{thetapm} gives the equivalent circuit,
\begin{equation} \label{Vdagcircuit}
\begin{split}
 \Qcircuit @C=1.2em @R=1.5em {                                              
           &\gate{\frac{1}{\sqrt{\pi}}V^{\dag}(\theta_a, \theta_b)} &\gate{D^{\dag}(\mu)} &\ar @{-} [dr(0.5)] \qw  &\\
                            &\qw                                   &\qw                  &\ar @{-} [ur(0.5)] \qw  & 
    } \; \raisebox{-1em}{,}
\end{split}
\end{equation}
where we define the operator [Eq.~\eqref{Vgate}]
    \begin{equation} \label{Vappendix}
        \op{V}(\theta_a, \theta_b) \coloneqq \op R(\theta_+ - \tfrac{\pi}{2}) \op S(\tan{\theta_-})\op R(\theta_+)\, ,
    \end{equation}
with $\theta_\pm = \frac{1}{2}(\theta_a \pm \theta_b)$ [Eq.~\eqref{measurementangles}]. Note that we have ignored an overall phase in this derivation, which results from combining the two displacements into~$\mu$.
    
Taking the Hermitian conjugate of the circuit in Eq.~\eqref{Vdagcircuit} gives a description of the two-mode entangled measurement in Eq.~\eqref{twomodemeasurement},
\begin{equation}\label{Vopcircuit}
\begin{split}
\centering
\resizebox{.8\columnwidth}{!}{
    \Qcircuit @C=1em @R=1.5em 
    {
	&&\lstick{\brasub{m_a}{p_{\theta_a}}} &\bsbal{1} &\qw &                      &&\ar @{-} [dl(0.5)] &\gate{D(\mu)} &\gate{\frac{1}{\sqrt{\pi}}V(\theta_a ,\theta_b)} & \qw \\
	&&\lstick{\brasub{m_b}{p_{\theta_b}}} &\qw       &\qw &\raisebox{2.5em}{$=$} &&\ar @{-} [ul(0.5)] &\qw           &\qw                                              & \qw \\
	}\, \raisebox{-1.5em}{.}
	}
	\end{split}
\end{equation}
This circuit contains the operators used in the main text and those in Refs.~\cite{Yokoyama2013, Ukai2010, Asavanant2020,Larsen2020}.
Note that when the Kraus state in the full teleportation gadget is nontrivial, then the outcome-dependent displacement $\op{D}(\mu)$ should be commuted past the teleported gate $\bounceEPR{\psi,\phi}{}$ in order to determine the necessary correction at the end of the gadget.

We present several variations of the circuit in Eq.~\eqref{Vopcircuit} by commuting the outcome-dependent displacement to different places among the other operators. For the case where the displacement acts first, we find
\begin{equation}\label{compareV}
\begin{split}
\centering
    \Qcircuit @C=1.2em @R=1.5em {
	& \ar @{-} [dl(0.5)] &  \gate{\frac{1}{\sqrt{\pi}}  V(\theta_a ,\theta_b)} &\gate{D(\mu') } &\qw \\
	& \ar @{-} [ul(0.5)] & \qw                           &\qw                                   &\qw \\
	}
\end{split}
\end{equation}
with amplitude
    \begin{align} \label{amplitudeprimed}
        \mu' &= \frac{ -m_a e^{-i \theta_b} + m_b e^{-i \theta_a } }{ \sin (\theta_a - \theta_b)}\, .
    \end{align}

We now discuss a different final form in order to complete the connection to the literature~\cite{Alexander2014,Alexander2018}. Pulling out a Fourier transform $F^\dagger = \op{R}(-\tfrac{\pi}{2})$ in Eq.~\eqref{Vappendix}, $\op{V}(\theta_a,\theta_b) = \op{F}^\dagger \op{V}''(\theta_a,\theta_b)$, we now consider the case where the displacement lies between the two other gates,
\begin{equation} \label{rafform}
    \begin{split}
    \centering
    \Qcircuit @C=1.2em @R=1.5em {
	& \ar @{-} [dl(0.5)]  & \gate{F^{\dag}} & \gate{ D(\mu'') } &  \gate{\frac{1}{\sqrt{\pi}} V''(\theta_a ,\theta_b)}  & \qw \\
	& \ar @{-} [ul(0.5)] & \qw                      & \qw      &\qw       & \qw \\
	}
	\end{split}
\end{equation}
and has amplitude
\begin{align}
    \mu'' &= \frac{-i m_a e^{i \theta_b} -i m_b e^{i \theta_a } }{ \sin (\theta_a - \theta_b)}\, .
\end{align}
This form is suitable for comparison with Refs.~\cite{Alexander2014,Alexander2018}, as $\op V''(\theta_a,\theta_b)$ and $\op D(\mu'')$ agree with those works up to one difference: their circuit does not contain the Fourier transform $\op F^{\dag}$.\footnote{Note that Ref.~\cite{Alexander2014} has an error in the denominator of the displacement operator: it appears in that reference as $\sin{\theta_-}$, while the correct quantity is $\sin(\theta_a-\theta_b)$.} This is because teleportation gadget in those works uses a two-mode canonical CV cluster state as its Kraus state, $\ket {\text{CVCS}} \coloneqq e^{i\op{q}\otimes \op{q}} \pket{0} \otimes \pket{0}$, which is related to the canonical EPR state by $\op{R}(\tfrac{\pi}{4})$ on each mode (or, equivalently, $\op{F}$ on either mode). Since $\op{q} = -\op{F} \op{p} \op{F}^\dagger$, this state can be represented with the following circuit identity,
\begin{align}
    \begin{split}
    \Qcircuit @C=1em @R=1.25em 
    {
 &\ctrl{1}  &\rstick{\pket{0}} \qw &&&&& \qw& \ctrl{1} & \qw & \rstick{\pket{0}} \qw\\
 &\control \qw &\rstick{\pket{0}} \qw &&&\raisebox{2em}{=}&&\gate{F}& \targ &\gate{F^{\dag}} & \rstick{ \pket{0} } \qw
		}
	\end{split}	
	\qquad \quad .
\end{align}
Acting with the Fourier transform gives $\op{F}^\dagger \pket{0} = \qket{0}$, and using Eq.~\eqref{EPRCX}, the resulting circuit is
\begin{equation} \label{eq:CVCSequiv}
\begin{split}
\raisebox{-1.2em}{$\ket {\text{CVCS}} \coloneqq {}$} \quad
    \Qcircuit @C=1em @R=1.1em 
    {
 &\ctrl{1}  &\rstick{\pket{0}} \qw &&&&& \qw & \ar @{-} [dr(0.5)] \qw & \\
 &\control \qw &\rstick{\pket{0}} \qw &&&\raisebox{2.3em}{=}&&       \gate{\frac{1}{\sqrt{2\pi}}F}  & \ar @{-} [ur(0.5)] \qw &
		}
    \; \raisebox{-1em}{,}
\end{split}
\end{equation}
recovering the well-known fact that teleportation with canonical CV cluster states is accompanied by a Fourier transform \cite{Menicucci2008}.
Thus, when using this Kraus state in conjunction with the measurement as described by Eq.~\eqref{rafform}, the two Fourier transform operators cancel in the final Kraus operator, $\op{F} \op{F}^\dagger = \op{I}$.

\begin{widetext}
\section{Glossary of circuit identities}
Here, we provide a succinct summary of the circuit identities in the main text. Note that all circuits are to be read right (input) to left (output).

\begin{itemize}
\setlength\itemsep{1.5em}

\item
Canonical maximally entangled EPR state, Eq.~\eqref{EPRCX}:
\begin{equation} 
       \begin{split}
 \Qcircuit @C=0.3cm @R=0.3cm {
         &&&&&&& \ctrl{1}  & \rstick{\pket{0}} \qw  &&&&                 &&&  \qw & \ar @{-} [dr(0.5)]\qw  &\\
         &&\ustick{\raisebox{.2em}{$\smash{\ket\EPR} \;\coloneqq$}}&&&&& \targ  & \rstick{\qket{0}}  \qw &&&& \ustick{\raisebox{.2em}{$=$}} &&& \gate{\frac{1}{\sqrt{2\pi}} I } & \ar @{-} [ur(0.5)] \qw  & 
    } 
       \end{split}
\end{equation}

\item Canonical two-mode CV cluster state, Eq.~\eqref{eq:CVCSequiv}:
\begin{equation}
\begin{split}
\raisebox{-1.3em}{$\ket {\text{CVCS}} \coloneqq {}$} \quad
    \Qcircuit @C=1em @R=1.1em 
    {
 &\ctrl{1}  &\rstick{\pket{0}} \qw &&&&& \qw & \ar @{-} [dr(0.5)] \qw & \\
 &\control \qw &\rstick{\pket{0}} \qw &&&\ustick{\raisebox{.2em}{$=$}}&&       \gate{\frac{1}{\sqrt{2\pi}}F} & \ar @{-} [ur(0.5)] \qw &
		}
\end{split}
\end{equation}

\item Square-lattice GKP Bell state, Eq.~\eqref{GKPBellpaircircuit}:
\begin{equation} \label{GKPBellpaircircuitapp}
\begin{split}
\raisebox{-1em}{$\ket {\Phi^+_\GKP} = {}$} \quad
    \Qcircuit @C=1em @R=1.2em 
    {
 &\bsbal{1}  &\rstick{\ket{\qunaught}} \qw &&&&& \qw & \ar @{-} [dr(0.5)] \qw & \\
 & \qw &\rstick{\ket{\qunaught}} \qw &&&\ustick{\raisebox{.2em}{$=$}}&&       \gate{ \frac{1}{\sqrt{2}}\,  \Pi_\GKP} & \ar @{-} [ur(0.5)] \qw &
		}
\end{split}
\end{equation}
with qunaught state $\ket{\qunaught}$, Eq.~\eqref{qunaught}, and GKP projector $\GKPproj$, Eq.~\eqref{GKPprojector}. 

\item Kraus state for a partial GKP projector ($p$ quadrature), Eq.~\eqref{GKPcircuitA}:
\begin{equation} \label{GKPcircuitAapp}
\begin{split}
    \Qcircuit @C=1em @R=1.1em 
    {
 &\bsbal{1}  &\rstick{\pket 0} \qw &&&&& \qw & \ar @{-} [dr(0.5)] \qw & \\
 & \qw &\rstick{\ket{\qunaught}} \qw &&&\ustick{\raisebox{.2em}{$=$}}&&       \gate{ 2^{\frac 1 4} \Sha_{\sqrt\pi}(\op p)} & \ar @{-} [ur(0.5)] \qw &
		}
\end{split}
\end{equation}

\item Kraus state for a partial GKP projector ($q$ quadrature), Eq.~\eqref{GKPcircuitB}:
\begin{equation} \label{GKPcircuitBapp}
\begin{split}
    \Qcircuit @C=1em @R=1.1em 
    {
 &\bsbal{1}  &\rstick{\ket \qunaught} \qw &&&&& \qw & \ar @{-} [dr(0.5)] \qw & \\
 & \qw &\rstick{\qket 0} \qw &&&\ustick{\raisebox{.2em}{$=$}}&&       \gate{ 2^{\frac 1 4} \Sha_{\sqrt\pi}(\op q)} & \ar @{-} [ur(0.5)] \qw &
		}
\end{split}
\end{equation}

\item Decomposition of the beam splitter, $\op B_{jk}$ in Eq.~\eqref{beamsplitterdef}, in terms of Gaussian gates, Eq.~\eqref{eq:BScircuitdecomp}, using notation introduced in Eq.~\eqref{BScircuit}:
    \begin{equation}
    \begin{split}
         \Qcircuit @C=1em @R=2.5em @! 
         {
         &	& \bsbal{1} & \qw \\
         & 	& \qw       & \qw
  		  }
\quad \raisebox{-1.3em}{$=$} \quad
    \Qcircuit @C=1em @R=1em {
         	& \ctrl{1} & \gate{S^\dag({\sqrt{2}})} & \targ &  \qw\\
         	& \targ & \gate{S(\sqrt{2})} & \ctrlo{-1} & \qw
  		  }
    \end{split} 
	\end{equation}

\item Bouncing operations from one mode to another through an EPR state, Eq.~\eqref{scrivauto:88}:
\begin{align} \label{eq:bounceapp}
    \begin{split}
    \Qcircuit @C=1em @R=1em {
    &\qw & \gate{O} & \qw[-1] \ar @{-} [dr(0.5)] & & & &\qw & \qw & \qw[-1] \ar @{-} [dr(0.5)] &\\
    &\qw & \qw & \qw[-1] \ar @{-} [ur(0.5)] & & \raisebox{2em}{$=$} & &\qw & \gate{O^{\tp}} & \qw[-1] \ar @{-} [ur(0.5)] & \, \\
    }
    \end{split}
\end{align}
with the transpose taken in the position basis.

\item State produced by mixing a t-momentum and s-position eigenstate on a beam splitter, Eq.~\eqref{entangleBS}:
  \begin{equation}
        \begin{split}
		 \Qcircuit @C=1em @R=1.5em {
         	& \bsbal{1} & \rstick{\pket{t}} \qw &&&&                     &&& \qw                                                               & \ar @{-} [dr(0.5)]\qw  &\\
         	& \qw       & \rstick{\qket{s}} \qw &&&& \raisebox{2em}{$=$} &&& \gate{\frac{1}{\sqrt{\pi}} D(s+it) } & \ar @{-} [ur(0.5)] \qw & 
  		  } \, 
  		\end{split}
	\end{equation}

\item Entangled two-mode measurement of rotated quadratures with outcomes $m_a$ and $m_b$, Eq.~\eqref{measurementcircuit}:
	\begin{equation}\label{measurementcircuitAppendix}
\begin{split}
    \Qcircuit @C=1em @R=1.5em 
    {
	&&\lstick{\brasub{m_a}{p_{\theta_a}}} &\bsbal{1} &\qw &                       &&\ar @{-} [dl(0.5)] &\gate{D(\mu) }  &\gate{ \frac{1}{\sqrt{\pi}} V(\theta_a ,\theta_b)} &\qw \\
	&&\lstick{\brasub{m_b}{p_{\theta_b}}} &\qw     &\qw   & \raisebox{2.5em}{$=$} &&\ar @{-} [ul(0.5)] &\qw                          &\qw                                                &\qw \\
	}
	\end{split}
\end{equation}
with Gaussian unitary $\op V(\theta_a ,\theta_b)$, Eq.~\eqref{Vgate}, and outcome-dependent displacement with amplitude $\mu$, Eq.~\eqref{mainmu}.

\item Beam splitter-entangled Kraus state for identical local damping on the ancillae, Eq.~\eqref{dampedancillacircuit}:
	\begin{equation} 
 \Qcircuit @C=0.25cm @R=0.3cm 
    {
         &\qw &\bsbal{1} &\qw &\gate{\dampgate} &\rstick{\ket{\psi}} \qw &&&&                              &&\qw              &\qw                                                      & \qw              & \ar @{-} [dr(0.5)] \qw  &\\
         &\qw &\qw       &\qw &\gate{\dampgate} &\rstick{\ket{\phi}} \qw &&&&\ustick{\raisebox{.5em}{$=$}} &&\gate{\dampgate} &\gate{ \frac{1}{\sqrt{\pi}} \bounceEPRgate{\psi, \phi}{}}& \gate{\dampgate} &  \ar @{-} [ur(0.5)] \qw  & 
    }
\end{equation}
for damping operator $\dampop{\beta}$ in Eq.~\eqref{damping}. This state can be normalized with factor $(\mathcal{N}_\phi \mathcal{N}_\psi)^{-1/2}$. 

\item Gate-teleportation gadget with damped ancillae comprising the Kraus state, Eq.~\eqref{genkrausdamp}:
    \begin{align}
    \begin{split}
    \scalebox{0.9}
    {
    \Qcircuit @C=1.25em @R=1em 
    {
	&\lstick{\brasub{m_a}{p_{\theta_a}}} &\bsbal{1} &\qw       &\qw              & \rstick{\text{(in)}} \qw[-1] &  \\
	&\lstick{\brasub{m_b}{p_{\theta_b}}} &\qw       &\bsbal{1} &\gate{\dampgate} & \rstick{\ket{\psi}} \qw \\
	&\lstick{\text{(out)}}	                     &\qw       &\qw       &\gate{\dampgate} & \rstick{ \ket{\phi} } \qw \\
	}
\qquad \raisebox{-2.3em}{=} \qquad	 
	   \raisebox{-2em}{ \Qcircuit @C=1em @R=1em {
& \lstick{\text{(out)}} & \gate{\dampgate}& \gate{\frac{1}{\sqrt{\pi}} \bounceEPRgate{\psi, \phi}{} } & \gate{\dampgate} & \gate{D(\mu)}) &\gate{\frac{1}{\sqrt{\pi}}V(\theta_a ,\theta_b)} &\rstick{\text{(in)}}\qw \\
	} \, }
	}
	\end{split}	
    \end{align}
 
\item Case (AB)---state produced by mixing two damped qunaught states, Eq.~\eqref{approxqunaught}, on a beam splitter, Eqs.~\eqref{GKPBellpaircircuit} and \eqref{caseAB}:
    \begin{equation}
	\begin{split}
 \Qcircuit @C=0.45cm @R=0.3cm 
    {
         &\bsbal{1} &\rstick{\ket{\bar{\qunaught}}} \qw &&                               &&\gate{ \mathcal{N}^{-1/2}_\qunaught\dampgate} &\bsbal{1} &\rstick{\ket{\qunaught}} \qw &\\
         &\qw       &\rstick{\ket{\bar{\qunaught}}} \qw && \ustick{\raisebox{.5em}{$=$}} &&\gate{ \mathcal{N}^{-1/2}_\qunaught \dampgate} &\qw        &\rstick{\ket{\qunaught}} \qw &      
    }
    \quad \raisebox{-1.5em}{=} \quad
    \Qcircuit @C=1em @R=1.8em 
    {
 & \qw &\qw & \qw & \ar @{-} [dr(0.5)] \qw & \\
 &\gate{ \mathcal{N}^{-1/2}_\qunaught \dampgate}&       \gate{ \frac{1}{\sqrt{2}} \Pi_\GKP} &\gate{ \mathcal{N}^{-1/2}_\qunaught\dampgate} & \ar @{-} [ur(0.5)] \qw &
		}
     \,
    \end{split}
    \end{equation}
The final form uses Eq.~\eqref{GKPBellpaircircuitapp} and \eqref{eq:bounceapp} above.

\item Case (A)---state produced by mixing a momentum-squeezed state, Eq.~\eqref{sqzstate}, and a damped qunaught state on a beam splitter, Eqs.~\eqref{caseAkraus} and \eqref{caseA}:
	\begin{equation}
	\begin{split}
 \Qcircuit @C=0.5cm @R=0.3cm 
    {
         & \bsbal{1} &\rstick{\pket{0; \zeta} }       \qw &&&                               &&\gate{ \mathcal{N}_\zeta^{-1/2} \dampgate}     &\bsbal{1} &\rstick{\pket{0}}       \qw &\\
         &\qw        &\rstick{\ket{\bar{\qunaught}} } \qw &&& \ustick{\raisebox{.5em}{$=$}} &&\gate{ \mathcal{N}_\qunaught^{-1/2} \dampgate} &\qw      &\rstick{\ket{\qunaught}}\qw & 
    }
    \quad \raisebox{-1.5em}{=} \quad
    \Qcircuit @C=1em @R=1.8em 
    {
 & \qw &\qw & \qw & \ar @{-} [dr(0.5)] \qw & \\
 &\gate{ \mathcal{N}^{-1/2}_\qunaught \dampgate}&       \gate{2^{\frac{1}{4}}\Sha_{\sqrt{\pi}}(\op{p})} &\gate{ \mathcal{N}^{-1/2}_\zeta \dampgate} & \ar @{-} [ur(0.5)] \qw &
		}
     \,
    \end{split}
    \end{equation}
The final form uses Eq.~\eqref{GKPcircuitAapp} and \eqref{eq:bounceapp} above.

\item Case (B)---state produced by mixing a damped qunaught state and a position-squeezed state, Eq.~\eqref{sqzstate}, on a beam splitter, Eqs.~\eqref{caseBkraus} and \eqref{caseB}:
	\begin{equation}
	\begin{split}
    \Qcircuit @C=0.5cm @R=0.3cm 
        {
         &\bsbal{1} &\rstick{\ket{\bar{\qunaught}}} \qw &&&                               &&\gate{ \mathcal{N}_\qunaught^{-1/2}\dampgate} &\bsbal{1} &\rstick{\ket{\qunaught}} \qw &\\
         &\qw        &\rstick{\qket{0; \zeta} }     \qw &&& \ustick{\raisebox{.5em}{$=$}} &&\gate{ \mathcal{N}_\zeta^{-1/2}\dampgate}     &\qw       &\rstick{\qket{0}}\qw & 
         }
         \quad \raisebox{-1.5em}{=} \quad
    \Qcircuit @C=1em @R=1.8em 
    {
 & \qw &\qw & \qw & \ar @{-} [dr(0.5)] \qw & \\
 &\gate{ \mathcal{N}^{-1/2}_\zeta \dampgate}&       \gate{2^{\frac{1}{4}} \Sha_{\sqrt{\pi}}(\op{q})} &\gate{ \mathcal{N}^{-1/2}_\qunaught\dampgate} & \ar @{-} [ur(0.5)] \qw &
		}
     \,
    \end{split}
    \end{equation}
The final form uses Eqs.~\eqref{GKPcircuitBapp} and \eqref{eq:bounceapp} above.
    
\end{itemize}

\end{widetext}

\bibliographystyle{IEEEtran} 
\bibliography{ReferencesMaster, rafref}
\end{document}